\documentclass{aa}  

\usepackage{graphicx}
\usepackage{txfonts}

\usepackage{bm}

\usepackage{algorithm}
\usepackage{algpseudocodex}

\usepackage{pifont}

\usepackage{booktabs}
\usepackage{multirow}
\usepackage{array}

\usepackage{subcaption}

\newcommand\norm[1]{\left\lVert#1\right\rVert}

\usepackage{hyperref}

\usepackage{stfloats}

\begin{document}

   \title{PolyCLEAN: Atomic Optimization for Super-Resolution Imaging \\
        and Uncertainty Estimation in Radio Interferometry}
    \titlerunning{PolyCLEAN: Atomic Optimization in Radio Interferometry}

    \author{Adrian Jarret \inst{1}
        \and
        Sepand Kashani\inst{1}
        \and
        Joan Rué-Queralt\inst{2}
        \and
        Paul Hurley\inst{3}\fnmsep\inst{4}
        \and \\
        Julien Fageot \inst{1}
        \and
        Matthieu Simeoni \inst{2}
        }

    \institute{Audiovisual Communications Laboratory, \'Ecole Polytechnique Fédérale de Lausanne, Switzerland\\
        \email{adrian.jarret@epfl.ch}
        \and
        Center for Imaging, \'Ecole Polytechnique Fédérale de Lausanne, Switzerland
        \and
        Centre For Research In Mathematics and Data Science, Western Sydney University, Australia
        \and
        International Centre for Neuromorphic Systems, Western Sydney University, Australia
        }

    \date{Submitted 3 June 2024; Revised 25 October 2024; Accepted <day month> 2024}

    \abstract
    {Imaging in radio interferometry amounts to solving an ill-posed noisy inverse problem, for which the most adopted algorithm is the original CLEAN algorithm and its variants. Alternative explicit optimization methods have gained increasing attention, as they demonstrate excellent reconstruction quality due to their ability to enforce Bayesian priors. Nowadays, the main limitation to their adoption lies in run-time speed. Additionally, uncertainty quantification is difficult for both CLEAN and convex optimization techniques.
    }
    {We address two issues for the adoption of convex optimization in radio interferometric imaging. First, a method for a fine resolution setup is proposed which scales naturally in terms of memory usage and reconstruction speed. Second, a new tool to localize a region of uncertainty is developed, paving the way for quantitative imaging in radio interferometry.
    }
    {The classical $\ell_1$ penalty is used to turn the inverse problem into a sparsity-promoting optimization. For efficient implementation, the so-called Frank-Wolfe algorithm is used together with a \textit{polyatomic} refinement. The algorithm naturally produces sparse images at each iteration, leveraged to reduce  memory and computational requirements. In that regard, PolyCLEAN reproduces the numerical behavior of CLEAN while guaranteeing that it solves the minimization problem of interest. Additionally, we introduce the concept of the \textit{dual certificate image}, which appears as a numerical byproduct of the Frank-Wolfe algorithm. This image is proposed as a tool for uncertainty quantification on the location of the recovered sources.
    }
    {PolyCLEAN demonstrates good scalability performance, in particular for fine-resolution grids. On simulations, the Python-based implementation is competitive with the fast numerically-optimized CLEAN solver. This acceleration does not affect image reconstruction quality: PolyCLEAN images are consistent with CLEAN-obtained ones for both point sources and diffuse emission recovery. We also highlight PolyCLEAN reconstruction capabilities on observed radio measurements.
    }
    {PolyCLEAN can be considered as an alternative to CLEAN in the radio interferometric imaging pipeline, as it enables the use of Bayesian priors without impacting the scalability and numerical performance of the imaging method.
    }


    \maketitle

%
\section{Introduction} \label{sec:intro}

Reconstructing an image in radio interferometry amounts to solving an inverse problem, the dimensions of which are determined by the number of interferometer baselines and the number of pixels used to cover the field of view. More measurements and longer baselines enhance the quality of reconstructed images, both in terms of sensitivity and resolution, but necessarily increase complexity. In the era of the Square Kilometer Array, the next generation radio interferometer that aims to contain hundreds of stations and thousands of antennas \citep{Scaife_2020}, this increased complexity causes considerable numerical challenges, whether it concerns data measurement, transfer, storage or processing.

In particular, it is crucial for the \textit{imaging algorithms} to be fast, reliable and scalable.
The most simple way to obtain a meaningful image consists of constructing the \textit{dirty image} with elementary inversion of the observed visibility. Although cheap to produce -- it only requires a single call to the adjoint of the measurement operator -- the dirty image presents reconstruction artifacts due to the partial coverage of visibility space and is thus rarely satisfactory.
The current imaging methods are more advanced and significantly improve the quality of the reconstruction. They can be classified into two families: adhering either to the CLEAN or the Bayesian paradigm. These methods introduce a prior image model, usually built around a notion of sparsity in the image, to account for ill-posedness. The two families however differ in the manner of enforcing the model.

\subsection{The CLEAN realm}
CLEAN traces its origins back to 1974 with the Högbom algorithm \citep{Hogbom_1974}, predating sparse reconstruction methods developed in the signal processing community by about two decades -- see the historical review in \cite[Notes of Chapter~1]{Foucart_Rauhut_2013} and references therein. As one of the first sparsity-promoting methods, this pioneering algorithm has had huge impacts in radio astronomy and beyond \citep{Cornwell_2009}. CLEAN creates a model for the reconstructed image by iteratively placing elementary components called atoms. These atoms are point sources in the case of the original Högbom algorithm, or building elements with a spatial extension in the case of multi-scale reconstruction for diffuse emissions, as in Asp-CLEAN \citep{Bhatnagar_Cornwell_2004} and MS-CLEAN \citep{Cornwell_2008}. Many developments have followed, creating a wide landscape of CLEAN-based image reconstruction techniques, improving the image model and numerical performance \citep{clark1980, schwab1984, offringa2014, Offringa_Smirnov_2017, muller2023a}. The family of CLEAN algorithms is based on a prior model that assumes that the image to recover is sparse; that is, it can be accurately approximated with a limited number of atoms. This sparse prior model is enforced empirically along the reconstruction by controlling the number of iterations.

CLEAN algorithms are considered fast and scalable due to the low complexity of the operations performed. The long-term development efforts proposed since the first creation of CLEAN have resulted in mature software, with fast numerically optimized implementations available  --  see \texttt{WSCLEAN} \citep{offringa2014}. This maturity is one of the reasons CLEAN-based methods remain predominant in radio astronomy applications today.

\subsection{The Bayesian paradigm}
The other grand family of imaging methods used in radio interferometry relies on the \textit{Bayesian statistical inference} framework. There, the prior image model is explicitly described in a mathematical manner using a probability distribution. Different image estimators then lead to various reconstruction algorithms covering a wide range of scalability properties. A trade-off between the completeness of the information recovered and numerical performance can usually be readily obtained. On one end of the spectrum are methods that estimate the full posterior distribution of the data, such as \cite{junklewitz2016, Cai_Pereyra_McEwen_2018a, arras2021}. Although these techniques recover extensive and precise information, they are numerically heavy, requiring the use of Monte-Carlo Markov Chains or Langevin sampling algorithms. For this reason, they are difficult to scale when the dimensions of the problem increase. On the other end of the information-scalability trade-off are the \textit{Maximum A Posteriori} estimators, proposed to the radio interferometry more than a decade ago \citep{Wiaux_Jacques_Puy_Scaife_Vandergheynst_2009, Li_Cornwell_Hoog_2011}. These estimators turn out to be more practical as they generally amount to solving a sparsity-promoting optimization problem. Although on the whole numerically heavier to implement than CLEAN-based techniques, run-time can still be reasonable. They are able to enforce complex priors, matching the versatility of CLEAN by recovering both point sources and diffuse emissions, and have demonstrated excellent reconstruction quality \citep{Schwardt_2012, Carrillo_McEwen_Wiaux_2012, Garsden_2015, chael2016, Akiyama_2017,Abdulaziz_Dabbech_Wiaux_2019,muller2023}.
The optimization methods however generally rely on proximal algorithms, which pose other concerns and thus have limited the overall adoption of these methods. On the one hand, proximal algorithms can be difficult to scale to large datasets. Indeed, they compute dense gradient steps at each iteration, which is not in line with the end goal of structurally sparse solutions. This mismatch leads to numerical overheads resulting in longer and less precise convergence. Some promising efforts have been made to scale out these algorithms, for instance with the faceted distributed algorithm in \cite{Thouvenin_2021}. On the other hand, most of the current imaging pipeline, including the calibration procedure, are built around atomic CLEAN-like strategies. Additional effort is thus required to integrate proximal algorithms, with a different algorithmic structure, into the calibration process.

It should be noted that learning approaches have also been proposed for radio interferometric imaging, including end-to-end learning \citep{schmidt2022}, reconstruction in the image domain \citep{aghabiglou2023a} and plug-and-play algorithms \citep{terris2022}. These approaches essentially learn a prior image model from training data. Contrary to the aforementioned families, this model is neither explicit nor fully understood. Thus, trust issues may arise when analyzing reconstructions. Deep learning models are indeed known to sometimes create artifacts, referred to as \textit{hallucinations} in the machine learning community (see the discussion in \cite{gottschling2020troublesome}). As in other computational imaging applications, the learning methods offer exciting opportunities for radio interferometric imaging, together with their usual limitations. The inference task is usually performed quickly and in a scalable distributed manner, but the learning phase may be more problematic, both in terms of time and computing power. Due to the specificity of the learning methods, namely that each application needs a bespoke trained model, learning methods have not been considered in our scalability study.

\subsection{Image resolution as a quality-efficiency trade-off}
Irrespective of the algorithmic framework, reconstruction time and memory requirements are significantly impacted by the choice of pixel size. To cover the same field of view, a higher resolution uses smaller pixels and thus increases image size, degrading numerical performance. In radio interferometric imaging, the \textit{model image} produced is generally convolved with the CLEAN beam, whose angular resolution corresponds to the main lobe of the point-spread function of the interferometer. For this reason, it is common to chose a coarse pixel size, at about the nominal resolution of the interferometer (defined by the diffraction limit). However, the above-mentioned sparsity-based methods, including CLEAN, actually perform nonlinear reconstruction and thus are able to recover information at a scale finer then this nominal resolution. Super-resolved reconstruction techniques have then been proposed, both with optimization methods \citep{honma2014,dabbech2018,Akiyama_2017} as well as with CLEAN \citep{chael2016, Fried_1995}. In that regard, it is relevant to develop methods that employ finer pixels than the nominal resolution, as they allow for the recovery of more accurate images from the same measurement set, at the cost of more compute time. Pixel dimension should ideally be set to the smallest recoverable scale. To take full advantage of this finer resolution, the model image should either be convolved with a representative beam sharper than the CLEAN beam or presented without convolution at all.

\subsection{Our contributions}
PolyCLEAN, the algorithmic framework presented, takes advantage of the strengths of both CLEAN and proximal optimization families to perform scalable and fast imaging in a sub-nominal resolution setup. PolyCLEAN is based on an optimization problem and has been developed with a particular attention to numerical performance. In this regard, PolyCLEAN represents a step towards a more practical and trustful use of the optimization methods in radio interferometry imaging pipelines. The contributions are twofold. First, a Frank-Wolfe algorithm is used to solve the optimization problem of interest. The algorithm applies an atomic paradigm, in a CLEAN-like manner, which significantly differs from the proximal methods usually employed. This algorithmic structure, based on the use of sparse iterates, considerably improves the scalability of the reconstruction method. It combines a recent \textit{Polyatomic} variant of the classical Frank-Wolfe algorithm \citep{Jarret_Fageot_Simeoni_2022} with a sparsity-aware implementation of the forward operator of radio interferometry \citep{kashani2023}. Second, the \textit{dual certificate image} comes as a byproduct of the optimization problem and can be seen as a step towards quantitative imaging in radio interferometry (uncertainty and error quantification).

Section~\ref{sec:model} presents the astronomical data model, a summary captured from existing literature in radio interferometry to provide a comprehensive understanding of PolyCLEAN. Section~\ref{sec:clean} describes the CLEAN algorithm using the notation from the preceding section, with emphasis on the advantages and disadvantages of the algorithm. In a similar fashion, Sect.~\ref{sec:optim} introduces the mathematical statement of optimization methods through the prism of Bayesian inference, with a particular focus on the LASSO problem used as the recovery model for PolyCLEAN. This section also mathematically introduces the dual certificate image, which is a direct consequence of the convex analysis of the LASSO problem. This selected overview of radio interferometric imaging knowledge allows us to introduce unified notation and lays the groundwork for the detailed presentation of PolyCLEAN in Sect.~\ref{sec:pclean}. Section~\ref{sec:exp} reports on the numerical experiments to demonstrate the scalability of PolyCLEAN as well as its reconstruction capabilities, both on simulated and observed measurement sets.

\section{Inteferometric data model} \label{sec:model}

Measurement equations model the data acquisition in radio interferometry. We first recall a physical measurement equation in \ref{sec:physical-model}, then derive a discretized version of this equation in \ref{sec:digital-model}, leading to the inverse problem at the heart of digital imaging in radio astronomy.

\subsection{Physical model with a continuous sky}
\label{sec:physical-model}

In radio interferometry, the \textit{power flux density}, measured in Jansky (Jy), is the power of the received electric field. Sky images are produced by estimating the \textit{intensity} or \textit{brightness} which corresponds to the power flux density per unit solid angle. It is represented by a nonnegative function $I$ defined over the celestial sphere $\mathbb{S}^2$ \citep[Sect.~1.2.1]{Thompson_Moran_Swenson_2017}.

For simplicity, equations throughout assume calibrated data corrected from the perturbations of both the instrumental effects and the propagation effect \cite[Section~5.1]{vanderVeen_Wijnholds_Sardarabadi_2019}. 

\paragraph{Image model}
The \textit{point source} model assumes celestial objects emit from a single point without any spatial extension. In this model, the brightness function $I$ is represented as a sum of Dirac impulses $\delta$ where the weight is the intensity of the source (in Jansky), so that the sky image can be modelled as 
\begin{equation*}
    I (\ell, m) = \sum_i \alpha_i \delta(\ell - \ell_i, m - m_i)
\end{equation*}
for some weights $\alpha_i > 0$ and source locations $(\ell_i, m_i) \in [-1, 1]$. The resulting image is sparse (mostly empty with a limited number of nonzero-valued locations).

More general models include \textit{diffuse emissions}, which are cosmic objects producing emissions over a spatial extent. In this case, images are generally non-sparse.  Although point sources form a simpler model and are relevant for large fields of view, it is often necessary to consider extended sources as soon as the field of view is zoomed-in and narrower.

\paragraph{Measurement equation}
The brightness distribution is sampled by means of an interferometer, which is an array of radio telescopes located on the Earth's surface. The action of the interferometer is modelled by the following linear measurement equation
\begin{equation}
    \mathbf{V} = \Phi (I) + \bm{\varepsilon},
    \label{eq:measurements}
\end{equation}
in which $\mathbf{V} \in \mathbb{C}^L$ is the vector of the \textit{visibility measurements} and the linear functional $\Phi$ models the action of the interferometer. $\bm{\varepsilon} \in \mathbb{C}^L$ is the realization of a complex-valued random variable in order to account for instrumental noise, which will be characterized later on. 

Imaging is performed by solving the inverse problem \eqref{eq:measurements} to find a solution $I_\mathrm{sol}$ that can be seen as an approximation of the actual sky.  The measurement equation serves as the guiding principle of the imaging process.

\paragraph{Forward operator}
The forward operator $\Phi$ samples the \textit{continuous visibility function} $\mathcal{V}$ at the location of the baselines of the interferometer $(\mathbf{p}_k)_{k=1, \dots, L}$. These baselines are defined as the spatial difference between the receiving antennas and given in units of the observed wavelength\footnote{Note that this equation assumes the observation bandwidth to be small compared to the center frequency of observation.}, while the visibility function has the appearance of a 3D Fourier Transform with integration performed over a sphere:
\begin{equation*}
    \mathcal{V}(\mathbf{p}) = \iint_{\mathbb{S}^2} I(\mathbf{r}) e^{-\mathrm{j} 2 \pi \langle \mathbf{r}, \mathbf{p} \rangle} \mathrm{d}\mathbf{r}, \quad \mathbf{p} \in \mathbb{R}^3.
\end{equation*}
 We obtain the general radio interferometry equation\footnote{When $I$ contains point sources the notation of the integrals is mathematically incorrect and is to be understood as $\iint_{\mathbb{S}^2} \delta(\mathbf{r}) \mathrm{d}\mathbf{r} = \iint_{\mathbb{S}^2} \mathrm{d}\delta(\mathbf{r})$. } \cite[Chapter 3]{Thompson_Moran_Swenson_2017}
\begin{equation}
    \Phi(I)[k] = \mathcal{V}(\mathbf{p}_k) = \iint_{\mathbb{S}^2} I(\mathbf{r}) e^{-\mathrm{j} 2 \pi \langle \mathbf{r}, \mathbf{p}_k \rangle} \mathrm{d}\mathbf{r}
    \label{eq:phi-sphere}
\end{equation}
for $k \in \{1, \ ...\,, L\}$.

{It is useful to rewrite equation \eqref{eq:phi-sphere} using the tangent plane coordinates around the direction of observation. Let us introduce the \textit{direction cosine coordinates} $\mathbf{r} = (\ell, m, n(\ell, m)) \in \mathbb{S}^2$ where $\ell, m \in [-1, 1]$ and $n(\ell, m) = \sqrt{1 - \ell^2 - m^2}$. The measurement equation \eqref{eq:phi-sphere} then becomes
\begin{equation}
    \Phi(I)[k] = \iint_{[-1, 1]^2} \frac{I(\ell, m)}{n(\ell, m)} e^{-\mathrm{j} 2 \pi (u_k\ell + v_km)} \mathcal{W}(\ell, m; w_k) \mathrm{d}\ell\mathrm{d}m
    \label{eq:phi-plane}
\end{equation}
for $k \in \{1, \ ...\,, L\}$, 
with
$$\mathcal{W}(\ell, m; w) = e^{-\mathrm{j}2\pi w(n(\ell, m) - 1)}$$
and $(u_k, v_k, w_k)$ as the wavelength-normalized ground coordinates of the baseline $\mathbf{p}_k$.}

This forward operator corresponds to the \textit{natural weighting} of the visibilities: each measurement accounts for the same weight in the reconstruction procedure. Using \textit{uniform} or \textit{robust} weighting schemes is not relevant with our formalism as the inverse problem is solved from the measurement equation and is independent of manual engineering of the PSF (see \cite{Rau_Bhatnagar_Voronkov_Cornwell_2009,Onose_Dabbech_Wiaux_2017} for more details on the weighting schemes).

\paragraph{Noise model}
Although the observation noise can be modeled in a more accurate manner (see for instance \mbox{\cite{Simeoni_Hurley_2021}}), in this manuscript we consider a simplified noise model, as standard in radio interferometry. We assume the additive model of equation \eqref{eq:measurements} with $\bm{\varepsilon}$ distributed according to a complex-valued Gaussian white noise.

\begin{remark}[Van Cittert-Zernike]
    When the field of view is narrow (small values of $\ell$ and $m$) or when the baselines $\mathbf{p}_k$ are considered co-planar (coordinate $w=0$), equation \eqref{eq:phi-plane} reduces to the well known Van Cittert-Zernike equation \cite[Chapter 15]{Thompson_Moran_Swenson_2017}. However, for larger field of views, it is crucial to consider the $\mathcal{W}$-term in the definition of the visibility function.
\end{remark}

\begin{remark}[Well-posedness and field of view]
    Note that equations \eqref{eq:phi-sphere} and \eqref{eq:phi-plane} are valid as long as the integrals are well-defined. In practice, the radio antennas have a limited \textit{collection area}, which induces an antenna \textit{primary beam} of sensitivity. The beam pattern is maximum towards the zenith and rapidly decays with the angle of observation. A consequence is that antennas naturally limit the width of observation to a limited surface of the celestial sphere. To simplify notations and without loss of generality, it is assumed that the sky images $I$ considered in the equations are  supported on the observable field of view of the antennas, centered around the direction of observation. The integrals of \eqref{eq:phi-sphere} and \eqref{eq:phi-plane} then sum over a compact support and are well-posed.
\end{remark}

The physical inverse problem defined by \eqref{eq:measurements} deals with sky images defined on a continuous space domain. Some methods have been proposed to directly find continuous-domain solutions without resorting to explicit discretization -- for example the Bluebild algorithm \citep{Kashani_2017} that splits radio interferometric images into different level sets of energy, or LEAP \citep{Pan_Simeoni_Hurley_Blu_Vetterli_2017} that aims at placing point sources at arbitrary locations in a grid-free manner. Other options are available for working on the continuous-domain, for instance by using a grid-based dictionary reconstruction, where the dictionary atoms are continuous functions but the locations of the center of these elements lie on a discrete grid (see examples of such techniques in \cite{simeoni2020}). In this work, we classically choose to pixelize the sky images to obtain a fully discrete, finite-dimensional inverse problem. The next section describes the discretization scheme employed.

\subsection{Measurement equation for digital images}
\label{sec:digital-model}
In practical applications, it is common to reconstruct a 2D raster image of a given portion of the sky. The field of view coordinates $(\ell, m) \in [-1, 1]^2$ in equation \eqref{eq:phi-plane} are usually discretized with a uniform fine grid  $(\ell_i, m_j)_{i, j = 1, \dots, n}$ of size $N = n \times n$, and the sky image is seen as a real-valued square image represented as a matrix $\mathbf{I} \in \mathbb{R}^{n \times n} = \mathbb{R}^{N}$. The value of a pixel sums the contributions of all the sources that lie on the area it covers, integrating both point and extended sources, divided by the angle area covered by a pixel. Each pixel is a brightness or intensity measure, whose value is in Jansky per pixel (Jy.pix$^{-1}$). With this model in hand, an approximate measurement equation is derived from \eqref{eq:phi-plane} using the finite sums
\begin{equation}
    \bm{\Phi}(\mathbf{I})[k] = \sum_{i=1}^n \sum_{j=1}^n {\frac{\mathbf{I}[i,j]}{n(\ell_i, m_j)} e^{-\mathrm{j} 2 \pi (u_k\ell_i + v_km_j)} \mathcal{W}(\ell_i, m_j; w_k)},
\label{eq:phi-discrete}
\end{equation}
for $k \in \{1, \ ...\,, L\}$. Here, $\bm{\Phi}$ is the linear analysis operator responsible for the \textit{degridding} operation.

The numerical imaging methods are then based on the discrete counterpart of equation \eqref{eq:measurements}, that we write using the above-defined linear measurement operator $\bm{\Phi}: \mathbb{R}^{n \times n} \to \mathbb{C}^L$
\begin{equation}
    \mathbf{V} = \bm{\Phi} (\mathbf{I}) + \bm{\varepsilon}.
    \label{eq:discrete-measurements}
\end{equation}
This discrete model is classical in radio interferometry. The forward operator $\mathbf{\Phi}$ is usually defined as a sequence of linear transformations involving the discretization of the images over a cartesian grid and a Fourier transform \citep{Wiaux_Jacques_Puy_Scaife_Vandergheynst_2009,Rau_Bhatnagar_Voronkov_Cornwell_2009}. 

As it will be useful later on, we also state the expression of the adjoint operator $\bm{\Phi}^*: \mathbb{C}^L \to \mathbb{R}^{n \times n}$. For any complex-valued measurement vector $\mathbf{h} \in \mathbb{C}^L$, we have
\begin{equation}
    \bm{\Phi}^*(\mathbf{h})[i, j] = \mathfrak{Re}\left\{ \sum_{k=1}^L { \frac{h_k}{n(\ell_i, m_j)} e^{\mathrm{j}2\pi  (u_k \ell_i + v_k m_j)} \mathcal{W}^*(\ell_i, m_j; w_k) } \right\}, 
    \label{eq:adjoint-phi}
\end{equation}
for $i, j \in \{1, \ ...\,, n\}$, where $\mathfrak{Re}$ is the real part of a complex number. The operator $\bm{\Phi}^*$ handles the \textit{gridding} of the visibility: it performs the direct synthesis of an image based on the measurements. Under natural weighting the dirty image is computed as $\bm{\Phi}^* \mathbf{V}$, where $\mathbf{V}$ is the vector of measured visibilities. Note that \eqref{eq:phi-discrete} and \eqref{eq:adjoint-phi} can be seen as instances of a 2D non-uniform Fourier transform ( 3D if the $\mathcal{W}$-term needs to be considered), which allows for a fast and scalable implementation to be developed (more details in Sect.~ \ref{sec:nufft}).

In the remainder of this article, we will stick with the discrete formulation of the sky image and the associated discrete forward operator. 

\begin{remark}[Cartesian grid discretization]
    As a mathematical object, the continuous-domain sky image $I(\ell, m)$ lives on a given area on the surface of the sphere $\mathbb{S}^2$. In order to represent it digitally, with a finite set of values, we need to use a tesselation of the support of the image. The classical choice of a discrete Cartesian grid -- deployed here -- is motivated by the use of the FFT algorithm in order to evaluate the forward operator, but other tesselations could be considered. The main limitation of a uniform Cartesian grid is that it introduces distortion effects due to the surface of the sphere represented by each pixel not being uniform. It leads to aberrations that are particularly visible towards the borders of the image for large fields of view.
\end{remark}

Similar to its continuous counterpart, the discrete inverse problem \eqref{eq:discrete-measurements} is still ill-posed and corrupted by noise. It is usually assumed that the interesting solutions respect some properties such as positivity or sparsity, that help regularize the problem. It results in a prior model over the search space that may be explicit or not. Several methods are then available to identify a candidate solution using different ways to enforce their respective prior. In the following section, we express the classical CLEAN algorithm in the notation of this section.

\section{Imaging with CLEAN} \label{sec:clean}

The CLEAN algorithm remains extremely popular in the radio astronomy community, with new versions and refinements being regularly proposed. From a signal processing point of view, CLEAN is an instance of a \textit{Matching Pursuit} algorithm \citep{Lannes_1997,Mallat_Zhifeng_Zhang_1993}, which it predates by two decades. It fits a parametric signal model to the radio interferometry measurement equation while enforcing sparsity in the reconstruction coefficients.

\paragraph{Parametric model}
In the initial algorithm, referred to as \textit{Högbom CLEAN} \citep{Hogbom_1974}, the unknown discrete image is assumed to be a collection of point sources. This amounts to looking for a solution of the form
\begin{equation*}
    \mathbf{I}_{\mathrm{sol}} = \sum_k \alpha_k \bm{\delta}_{(i_k, j_k)} \in \mathbb{R}^{n \times n},
    \tag{PS}
    \label{eq:ps}
\end{equation*}
with $\bm{\delta}_{(i,j)}$ being an almost empty image with only the pixel $(i,j)$ at value $1$. The number of point sources in the image is assumed to be small compared to the dimensions of the image, which results in few nonzero $\alpha_k$ coefficients and a sparse solution overall.

Later versions of CLEAN considered images expressed as combinations of elements from redundant reconstruction dictionaries, leading to more complex image models. For instance, \textit{Multi-Scale CLEAN} (MS-CLEAN, \cite{Cornwell_2008}) uses a dictionary of isotropic reconstruction atoms with different width
\begin{equation*}
    \mathbf{I}_{\mathrm{sol}} = \sum_k \alpha_k \bm{g}_{(i_k, j_k)}^{\sigma_k},
    \tag{MS}
\end{equation*}
with $\bm{g}_{(i, j)}^{\sigma}$ being a discretized kernel with center on the pixel $(i,j)$ parameterized with a spatial extension $\sigma$. For instance, $\bm{g}_{(i, j)}^{\sigma}$ can be a Gaussian kernel, but other functions have also been used such as the prolate spheroidal functions (which have the advantage of being compactly supported). In such a model, the sparsity is still enforced on the coefficients $\alpha_k$, leading to a solution image that is usually not sparse in the image domain.

\paragraph{The algorithm}
Matching pursuit algorithms construct images by iteratively searching for the dictionary element having the highest correlation with the residual measurements. This dictionary element is then added to the current estimate of the solution, scaled by a multiplicative gain factor. The reconstruction is stopped when the current iterate is considered good enough, for example when the residual noise is smaller than a threshold or after a given number of iterations \citep{offringa2014}.

For the point source image model \eqref{eq:ps}, the reference CLEAN algorithm is \textit{Cotton-Schwab CLEAN} \citep{schwab1984} that reads as Algorithm \ref{alg:clean}, where the reconstruction parameters are the number of iterations $k_{max}$ and the gain $\alpha$.

\begin{algorithm}
\caption{Cotton-Schwab CLEAN (Major cycles only)}
\begin{algorithmic}
\State \textbf{Parameters :} $k_\mathrm{max}$ (iterations), $\alpha > 0$ (gain) \\
\State \textbf{Initialisation :} $\mathbf{I}^{(0)} = \mathbf{0}$, $\mathbf{I}_D = \mathbf{\Phi}^* \mathbf{V}$\\
\For {$k=1, 2,  \cdots, k_\mathrm{max}$}
    \State 1. Compute the dirty residual: $\mathbf{I}_R^{(k)} = \mathbf{I}_D - \mathbf{\Phi}^*\mathbf{\Phi} \mathbf{I}^{(k-1)}$
    \State 2. Find the location of the next reconstructed source: $$s^{(k)} = \underset{(i, j)}{\arg\max} {\left\lvert \mathbf{I}_R^{(k)}[i, j] \right\rvert} $$
    \State 3. Update the iterate: $\mathbf{I}^{(k)} = \mathbf{I}^{(k-1)} + \alpha (\max \mathbf{I}_R^{(k)}) \bm{\delta}_{s^{(k)}}$
\EndFor
\\
\State \textbf{Output:}
\State \qquad Postprocess $\mathbf{I}^{(k)}$ (convolution with synthetic beam, add residual image)
\end{algorithmic}
\label{alg:clean}
\end{algorithm}

\paragraph{Minor cycles}
The most demanding step in terms of memory and computation power is the application of the forward and adjoint operators in step 1 to compute the dirty residual image. To accelerate this step, the composed operator $\mathbf{\Phi}^*\mathbf{\Phi}$ is approximated as a convolution with the kernel $\mathbf{B_\mathbf{\Phi}}$, so-called the \textit{point spread function} (PSF) of $\mathbf{\Phi}$ and refered to as \textit{dirty beam}.
The cost of applying $\mathbf{\Phi}^*\mathbf{\Phi}$ is then reduced to simply performing a convolution, which is significantly faster than calling the exact operators but results in an approximation error. Indeed, observing a point source far from the center of the image induces a change of intensity in addition to the spatial shift. The error is small for narrow fields of view and increases with the width of the observation window. These approximate steps are called minor cycles, and are performed for most of the iterations of CLEAN. The major cycles, with the exact computation of the operator, are performed every few hundred (or thousand) of minor cycles to recompute the dirty residual with more accuracy.

\paragraph{Advantages and limitations}

The family of CLEAN-based algorithms has reached its popularity in radio astronomy as these methods are able to meet many of its imaging constraints. Later on, \textit{CLEAN} is used as a generic term to refer to any version, whether it is Cotton-Schwab for point sources or multiscale variants.
\begin{itemize}
    \item[\ding{51}] Thanks to the convolution-based minor cycles, CLEAN is fast  and performs few expensive calls to the measurement operator. In addition, the iterates admit a sparse representation, convenient to store and manipulate.
    \item[\ding{51}] CLEAN-like algorithms are flexible with respect to the image to reconstruct. The matching pursuit structure allows us to generalize the reconstruction to diffuse emissions using reconstruction dictionaries.
    \item[\ding{51}] CLEAN is well-suited for calibration. Indeed, the visibilities are calibrated by reconstructing a few bright and well-known point sources, which is conveniently handled with an atomic method like CLEAN.
    \item[\ding{51}] Last but not least, CLEAN-based algorithms have been present for a long time in the field of radio interferometric reconstruction. They thus benefit from extended development, including efficient and optimized implementations of the algorithms, and expertise from the researchers on how to use them to their full capacity.
\end{itemize}

\noindent In spite of these positive considerations, CLEAN-based algorithms have a few limitations, which we summarize as follows:
\begin{itemize}
    \item[\ding{56}] CLEAN, as a Matching Pursuit algorithm, does not perform denoising. The effect of noise in the measurement is only handled by enforcing sparsity in the solution.
    \item[\ding{56}] As a consequence and because Matching Pursuit tends to explain all the measurements by cancelling out the residuals, CLEAN needs to be stopped before complete convergence. The resulting solution is then highly sensitive to this arbitrary choice of stopping criterion.
    \item[\ding{56}] CLEAN often reconstructs regions with negative flux values that are physically unrealizable -- as for instance reported in \cite{arras2021}. This may result from an initial weight allocation that is too greedy or  tentative to explain the noise.
    \item[\ding{56}] Early termination of the algorithm prevents CLEAN from converging towards the final fit of Matching Pursuit. For this reason, the CLEAN solution can only be uncertainly interpreted as a sparse intermediate solution to a least-square problem, without any further guarantee of sparsity or recovery. See \cite{schwarz1978,solo2008, Lannes_1997} for more details on the convergence of CLEAN and \cite{locatello2018a} on Matching Pursuit.
    \item[\ding{56}] CLEAN does not provide uncertainty quantification on the solution.
\end{itemize}

In summary, the CLEAN algorithm and its refinements form a set of efficient tools for radio interferometric image reconstruction, designed and powered by decades of research  in the field. However, by essence, they face the same limitations as the Matching Pursuit algorithm for sparse recovery in its respective field of compressive sensing, that include robustness to noise and sparsity guarantees. In that regard, it is relevant to explore in radio interferometry the convex optimization methods that are generally used in place of Matching Pursuit algorithms when sparse reconstruction is involved. This line of research has already proven to be fruitful, with excellent results in terms of reconstruction quality (see for instance \cite{Thouvenin_Dabbech_Jiang_Abdulaziz_Thiran_Jackson_Wiaux_2023}). The next section presents the Bayesian inference framework that naturally leads to the optimization problems the PolyCLEAN method is based upon.

\section{Imaging with sparsity promoting Bayesian estimators} \label{sec:optim}

Bayesian theory provides aguably the most principled way to introduce prior information into a reconstruction problem. It relies on a user-defined prior probability distribution over the search space. Different estimators can then be used to reconstruct candidate solutions from the posterior distribution. For \textit{PolyCLEAN}, we use the sparsity-promoting \textit{Laplace distribution} prior and compute the \textit{Maximum A Posteriori} (MAP) estimator, which amounts to solving a LASSO optimization problem. Let us review how this problem appears and why it is well-suited to radio interferometric imaging.

\subsection{From Bayes posterior density to LASSO: MAP}

The Bayesian statistical inference framework makes the assumption that all components of our problem, namely the visibility measurements $\mathbf{V}$ and the source sky image $\mathbf{I}$, are realizations of random variables. A so-called \textit{posterior distribution} of the image $\mathbf{I}$ given the observations $\mathbf{V}$ is derived, denoted by $p(\mathbf{I}|\mathbf{V})$. Due to Bayes theorem, this quantity is proportional to the product of a \textit{likelihood} term $p(\mathbf{V}|\mathbf{I})$ and a \textit{prior distribution} term $p(\mathbf{I})$. The likelihood is derived from the measurement equation \eqref{eq:discrete-measurements}, involving the forward operator and the measurement noise model. Regarding a prior distribution, it is common to promote sparsity in the reconstruction by means of a Laplace distribution, expressed as $p(\mathbf{I}) \propto e^{-b \norm{\mathbf{I}}_1}$, where $b >0$ is a scale parameter. We refer the reader to \cite{Cai_Pereyra_McEwen_2018a} for a thorough and more complete presentation of the Bayesian inference framework applied to radio interferometric imaging.

The obtained posterior distribution contains the recovered information which makes it an extremely insightful scientific tool, though possibly difficult to manipulate. In practice, this posterior distribution may be either demanding to approximate numerically, or completely inaccessible. Estimators are then used to represent this distribution and produce reconstructions. In this work, we consider the \textit{maximum a posteriori} (MAP) estimator, which is defined as an image for which the posterior density reaches one of its maxima. The simplicity of this estimator makes its computation conceptually simple, as it amounts to solving the following LASSO optimization problem \citep{Cai_Pereyra_McEwen_2018b}:
\begin{equation}
    \underset{\mathbf{I} \in \mathbb{R}^N}{\arg\min} {\frac{1}{2} \lVert \mathbf{V} - \mathbf{\Phi}\mathbf{I} \rVert_2^2 + \lambda \norm{\mathbf{I}}_1},
    \label{eq:lasso}
\end{equation}
in which $\lambda>0$ is a regularization parameter.
This problem is known to produce sparse solutions, \textit{i.e.} solutions with only a few nonzero coefficients. For this reason, and as already documented in the literature \citep{Wiaux_Jacques_Puy_Scaife_Vandergheynst_2009,Li_Cornwell_Hoog_2011, Schwardt_2012, Garsden_2015}, the LASSO problem is relevant for the reconstruction of astronomical images containing point sources. Note that solutions are not necessarily unique, leading to potentially many images consistent with the observed visibilities.

\begin{remark}[Positivity constraint]
\label{remark:pos}
    Thanks to the flexibility of the Bayesian framework, a positivity constraint on the reconstructed images is easily introduced by changing the prior distribution. Although the use of such a positivity constraint may be considered optional for the CLEAN reconstructions, it usually improves the quality of the images obtained using optimization methods \cite[Sect.~4.5]{vanderVeen_Wijnholds_Sardarabadi_2019}.
\end{remark}

Parallel to Bayesian techniques, the LASSO problem is a classical tool in the compressed sensing community. The next section presents the grounding of LASSO in  optimization theory, which motivates its use as a robust and principled model for sparse image reconstruction.

\subsection{LASSO properties from optimization theory}
\label{sec:lassoprops}
The LASSO problem stands as a foundational method in the sparse reconstruction literature. Its name originated in the statistics community \citep{Tibshirani_1996}, also known as \textit{basis pursuit denoising} in compressive sensing \citep{Chen_Donoho_Saunders_2001}. Over the years, by means of convex optimization theory and dual analysis, many results have been derived to describe the solutions and the behavior of the problem, such as conditions for uniqueness \citep{Tibshirani_2013} or recovery guarantees (see the discussion in the introduction of \cite{duval2015} for an overview). Let us focus on two notions that help to understand the relevance of the LASSO in radio interferometric imaging.

\paragraph{Representer theorem}
The existence of sparse solutions to the LASSO problem is well-known (see for instance \cite{Elad_2010}).
Recent \textit{representer theorems} provide a new interpretation of this property by describing the solution set of many convex optimization problems \citep{Boyer_2019}. This new theory helps to build trust in the optimization-based reconstruction methods as their solutions are better understood and make  physical sense.
For the LASSO, the solution set is proven to be non-empty, compact and convex. The extreme points of this set are sparse elements, whose sparsity index is bounded by the number of measurements \citep{Unser_Fageot_Gupta_2016}. In a context of high undersampling, as is the case in radio interferometry, this representer theorem ensures the existence of sparse solutions to the LASSO.

\paragraph{Dual certificate}
Additionally, the LASSO problem benefits from the analysis of duality theory. For a given solution $\mathbf{I}_\mathrm{sol}$ of \eqref{eq:lasso}, the so-called \textit{dual certificate}\footnote{The terminology is common for continuous-domain problems but the definition still holds for finite-dimensional problems \citep{Duval_2017}.} $\bm{\mu}_\lambda$ is defined as
\begin{equation}
    \bm{\mu}_\lambda = \frac{1}{\lambda} \mathbf{\Phi}^*(\mathbf{V} - \mathbf{\Phi}\mathbf{I}_{\mathrm{sol}}).
    \label{eq:dual_certif}
\end{equation}
While there are in general multiple solutions to the LASSO they all share the same measurements, that is the \textit{fit} $\mathbf{\Phi I}_\mathrm{sol}$ is itself unique and independent of $\mathbf{I}_\mathrm{sol}$ \citep{Tibshirani_2013}. The dual certificate is thus unique and independent of the recovered solution.

This quantity can be used to extract information about all the possible solutions to the LASSO, given that at least one is known. Indeed, relying on the optimality conditions of the problem \citep[Proposition~1]{Duval_2017}, the dual certificate satisfies the two following properties for any solution $\mathbf{I}_\mathrm{sol}$:
\begin{align}
    \langle \bm{\mu}_\lambda, \mathbf{I}_{\mathrm{sol}} \rangle &= \norm{\mathbf{I}_{\mathrm{sol}}}_1, \label{eq:dual-saturation1}\\
    \norm{\bm{\mu}_\lambda}_\infty &\leq 1. \label{eq:dual-saturation2}
\end{align}
Combined together, these properties imply that the support of any solution is contained in the so-called \textit{saturation set} $\mathcal{D}_\lambda$ of the dual certificate, possibly empty, defined as the set of indices where $\bm{\mu}_\lambda$ reaches $1$ in absolute value
\begin{equation}
    \mathcal{D}_\lambda = \left\{ i \in \{1, \dots, N\} : |\bm{\mu}_\lambda[i]| = 1 \right\}.
    \label{eq:saturationset}
\end{equation}
This property is critical for a posteriori analysis of the reconstruction: given one solution to the LASSO, it is possible to compute the dual certificate and its saturation set and hence access the support of \textit{any other solution}.

This dual certificate takes the shape of an image, which makes it convenient to display, analyze and manipulate. We show an example in Sect.~\ref{sec:dual}. 

\subsection{Versatility of the LASSO for diffuse emissions}
\label{sec:mslasso}

The LASSO problem in its simple form reconstructs sparse images and thus is adapted for point source reconstruction. When diffuse emissions are involved images are not directly sparse, with lots of pixels expected to have a nonzero value. Nonetheless, the images may still be highly structured such that they can be represented by a limited number of components (and thus sparse in another domain). Such a representation can be obtained for example with wavelet transforms, which are known to produce sparse multi-scale component images. In particular, the \textit{Isotropic Undecimated Wavelet Transform} is considered well-suited for representing radio astronomical images \citep{starck2006}. Using a modified LASSO problem, many methods have been developed to recover images with a sparse decomposition in a multi-scale wavelet frame \citep{Carrillo_McEwen_Wiaux_2012, Garsden_2015}.

Another way to rely on sparsity for diffuse emission reconstruction is to assume that the sky image can be expressed (or approximated) with a small number of elements from a large dictionary of reconstruction atoms, which is again a form of sparsity. This is the approach used by MS-CLEAN, in which the dictionary elements are instances of \textit{prolate spheroidal functions} with different scales \citep{Cornwell_2008}.

Whether a dictionary of atoms or a wavelet frame is used, the image to recover is expressed as
\begin{equation}
    \mathbf{I} = \bm{\Psi\theta},
    \label{eq:params-synthesis}
\end{equation}
where $\bm{\Psi}$ is a linear synthesis operator that maps the decomposition coefficients $\bm{\theta} \in \mathbb{R}^M$ to an actual image $\mathbf{I} \in \mathbb{R}^N$. The sparsity hypothesis now applies to $\bm{\theta}$. In \texttt{WSCLEAN}, the reconstruction coefficients are determined with an iterative greedy strategy. With PolyCLEAN, we instead solve the following LASSO problem, known as \textit{synthesis form}:

\begin{equation}
        \underset{\mathbf{\theta} \in \mathbb{R}^{M}}{\arg \min} \frac{1}{2} \lVert \mathbf{V} - \mathbf{\Phi}\mathbf{\Psi}\bm{\theta} \rVert_2^2 + \lambda \norm{\bm{\theta}}_1.
    \label{eq:lasso-synthesis}
\end{equation}

\begin{remark}[Analysis form]
    It is possible to consider the problem in its so-called \textit{analysis form}. The optimization variable is then $\mathbf{I}$ and the synthesis operator goes within the regularization term as the adjoint operator $\norm{\mathbf{\Psi}^*\mathbf{I}}_1$. In the general case, the problems are not equivalent and the analysis form can lead to better results \citep{elad2006,tibshirani2011b,carrera2017}. In \cite{Cai_Pereyra_McEwen_2018b}, both analysis and synthesis forms demonstrated no major discernible difference. The PolyCLEAN framework relies on the synthesis form as it is more convenient to treat with a Frank-Wolfe algorithm (see Sect.~\ref{sec:pclean}).
\end{remark}

\subsection{Proximal solvers and their numerical limitations}

LASSO, similarly to other optimization techniques in radio interferometry, has a non-differentiable objective function which cannot be minimized by simple gradient steps. These non-differentiable problems are generally solved with iterative \textit{Forward-Backward} schemes, a class of algorithms that leverage the existence of a simple closed-form expression for the penalty term \citep{parikh2014}. In particular, the LASSO problem is usually solved with the \textit{Accelerated Proximal Gradient Descent} algorithm (APGD), also known as FISTA \citep{beck2009}. These proximal solvers are flexible and show fast convergence properties. For the LASSO, APGD is proven to have an optimal convergence speed in terms of number of iterations \citep{liang2022}.

However, for sparse reconstruction, especially in high dimensional settings, the proximal methods may suffer from scalability limitations when compared to atomic methods such as CLEAN \citep{Offringa_Smirnov_2017}. Although the end solution is sparse, the iterates are usually dense with very little zero-valued elements. This structural mismatch between iterative method and the nature of the solution might be prejudicial for two reasons: 1) the algorithm may require a large number of iterations to sparsify the iterate, leading to a slow convergence; and 2) the solver needs to store the iterates in a dense format along the iterations, increasing memory requirements and affecting scalability. 

In contrast, the PolyCLEAN method demonstrates an atomic CLEAN-like behavior, which allows sparse iterates to be maintained across iterations. In doing so, PolyCLEAN preserves the strengths of CLEAN presented in Sect.~\ref{sec:clean} while improving reconstruction thanks to the benefits of the optimization framework. The next section presents the algorithm and its scalability-oriented design.

\section{The PolyCLEAN pipeline as an atomic imaging method} \label{sec:pclean}

Let us summarize the three major bricks of the PolyCLEAN framework:
\begin{itemize}
    \item \textbf{LASSO ---} The imaging inverse problem \eqref{eq:discrete-measurements} is solved using the sparsity prior of the LASSO problem, namely using an $\ell_1$-norm regularizer. The dual certificate comes as a natural output of the LASSO and provides information on the localization of the reconstructed sources. PolyCLEAN can optionally enforce a positivity constraint on the image. PolyCLEAN also supports the LASSO problem formulated in synthesis form over a transformed domain, to perform sparse dictionary or sparse wavelet reconstruction (as described in Sect.~\ref{sec:mslasso}).
    \item \textbf{The Polyatomic Frank-Wolfe algorithm ---} The LASSO problem \eqref{eq:lasso} is minimized by the \textit{Polyatomic Frank-Wolfe} algorithm (PFW), recently developed by the authors \citep{Jarret_Fageot_Simeoni_2022}. This algorithm has been specifically designed to solve sparse inverse problems in large dimensional setups, which makes it relevant to radio astronomy applications.
    \item \textbf{A sparsity-aware implementation of the interferometric forward operator ---} Independently of the solving method, applying the forward operator $\mathbf{\Phi}$ is both time and memory consuming, as it is usually done via a (de-)gridding operation followed by a multidimensional fast Fourier transform. In PolyCLEAN we implement this operator using a \textit{Non-Uniform Fast Fourier Transform} (NUFFT) algorithm. We rely on the \texttt{HVOX} package to do so, a chunking technique for efficient computation of the NUFFT in Python \citep{kashani2023}. The sparse structure of both the images and the frequency observations are the key ingredients in making computations faster and reducing memory requirements of the forward and adjoint operators.
\end{itemize}
With this structure, the benefits of PolyCLEAN are twofold. On the one hand, thanks to its CLEAN-like atomic behavior, PolyCLEAN provides a fast and scalable algorithm to perform sparsity-inducing Bayesian imaging, improving upon the proximal solvers in terms of computation time and memory requirements. The Polyatomic Frank-Wolfe algorithm and the \texttt{HVOX}-based forward operator both independently enable  speed and scalability improvements. Their combination within PolyCLEAN results in an efficient and elegant way to build upon their respective computational strengths. 

On the other hand, as PolyCLEAN is based on a LASSO problem, it is intrinsically connected to the LASSO dual certificate. The powerful theoretical guarantees of this dual certificate -- identification of the solution support -- mean it can become an insightful tool for quantification of error in images. At each iteration, the Frank-Wolfe algorithm computes the \textit{empirical dual certificate} that converges toward the final certificate. Although the dual certificate is a common tool in convex optimization theory, to the best of our knowledge this is the first time it is deployed for application purposes in radio interferometry.

The next paragraphs present the Polyatomic Frank-Wolfe algorithm, the NUFFT operator and the analysis of the dual certificate in the context of radio interferometry.

\subsection{Solving with Polyatomic Frank-Wolfe}

The backbone of PolyCLEAN consists in applying the Polyatomic Frank-Wolfe algorithm (PFW) recently proposed in \cite{Jarret_Fageot_Simeoni_2022} to solve the LASSO problem of interest.

For the sake of simplicity we limit discussion in this section to the case of point source reconstruction. When diffuse emissions are present, the interferometric problem becomes the LASSO in synthesis form \eqref{eq:lasso-synthesis}, which has a modified forward operator.

\paragraph{Renewed interest in the Frank-Wolfe algorithms}

The original Frank-Wolfe algorithm is a classical convex optimization algorithm \citep{Frank_Wolfe_1956} that constructs a solution as a sum of atoms in a projection-free manner. This atomic behavior has been identified to produce a fast and scalable algorithm, particularly in the context of sparse reconstruction, leading to a revival of interest from the optimization community over the past decade \citep{jaggi2013, harchaoui2013}. The Frank-Wolfe algorithm now benefits from many variants \citep{jaggi2013, lacoste-julien2015}, and the connection has been drawn with other sparse reconstruction methods, including Matching Pursuit \citep{locatello2017}. In particular, when Frank-Wolfe is applied to the LASSO problem \citep{Jarret_Fageot_Simeoni_2022, harchaoui2013}, the algorithm demonstrates a striking numerical similarity with CLEAN. The reconstruction atoms are indeed point sources, identified by single pixel images. In that regard, Frank-Wolfe algorithms are good candidates to fit in a CLEAN numerical reconstruction pipeline, as they guarantee the convergence of the reconstruction towards a solution to the LASSO.

\paragraph{Benefits of PFW}
The polyatomic variation of Frank-Wolfe is an additional refinement that allows the algorithm to place many atoms at each iteration instead of only one at a time. This greatly reduces the solving time and is the key ingredient that brings a Frank-Wolfe algorithm up to speed with the proximal methods \citep{Jarret_Fageot_Simeoni_2022}. In a sense, this polyatomic strategy can be compared to the minor cycles in CLEAN. It is another way to populate the solution with many candidate atoms without evaluating the forward operator. PFW is proven to converge towards a minimum of the objective function of the LASSO. The name \textit{PolyCLEAN} arises naturally from a portmanteau of a polyatomic update with a CLEAN-like algorithm.

Compared to APGD, PFW is more scalable and faster in the specific context of high sparsity. Indeed, as the iterates of PFW are sparse, their memory usage is limited, whereas APGD has dense iterates due to the gradient steps. Regarding the reconstruction speed, the theoretical rate of PFW is slower than APGD. However, numerical applications demonstrate significantly faster convergence for PFW (see Sect.~\ref{sec:profile}). This is another consequence of the sparse iterates which allow the algorithm to quickly identify the few active degrees of freedom of a sparse solution.

Finally, it is noteworthy that, despite its acceleration, PFW is a reliable algorithm as it maintains convergence guarantees. As demonstrated in \cite{Jarret_Fageot_Simeoni_2022}, the sequence of iterates converges towards a minimum of the LASSO objective function.

\paragraph{The algorithmic steps}
The detailed steps are provided in Algorithm \ref{alg:pfw}. The procedure starts with an empty sky. Each iteration is composed of two steps. During the first, the algorithm identifies a set of candidate atoms located at the regions of maximum intensity of the residual image.  This differs from CLEAN, which only looks for a single atom placed at the maximum intensity of the residual image. These candidate atoms are stored in $\mathcal{S}_k$, the set of active indices. The parameter $\delta$ controls the intensity of the polyatomic behavior.

During the second step of each iteration, PolyCLEAN simultaneously evaluates the weight to associate to each new atom and adjusts the weights already attributed to the previously selected atoms. Moreover, the algorithm prunes the set of atoms, removing those which get attributed zero weight. This second step is handled by solving another LASSO problem, defined as the restriction of the initial LASSO problem with the constraint that the support of the solution is included in $\mathcal{S}_k$. This constraint significantly reduces the dimensionality of the problem so that this LASSO sub-problem can be solved efficiently both in terms of memory and runtime.

To further accelerate the second step, the solver of the subproblem is pre-emptively stopped with initially low accuracy. This early stop does not affect overall reconstruction quality, as the accuracy of the reweighting is increased over the iterations, and this simply results in a speedup (see \cite{Jarret_Fageot_Simeoni_2022} for more details). An instance of an APGD algorithm is used to perform this constrained minimization.

Finally, the algorithm stops if a user-defined stopping criterion is met. As PFW is guaranteed to converge towards a minimum of the LASSO objective function, we naturally use a criterion of small relative improvement of the objective function value between the iterations.

\begin{algorithm}
\caption{Polyatomic FW of quality $0 < \delta \leq 1$}
\label{alg:pfw}
\begin{algorithmic}
    \State \textbf{Initialize:} $\mathbf{I}_0 \gets 0, \mathcal{S}_0 \gets \emptyset, \Delta \gets (1 - \delta) \norm{\mathbf{\Phi}^*\mathbf{V}}_\infty$ \\
    
    \For{$k=0, 1, 2, \dots$}
        \State Dirty residual: $\bm{\eta}_k \gets \mathbf{\Phi}^*\left( \mathbf{V} - \mathbf{\Phi}(\mathbf{I}_k) \right)$
        \vspace{6pt}
        \State 1.a. Polyatomic exploration:
        \State \qquad $\mathcal{I}_{k+1} = \left\{ 1 \leq j \leq N : |\bm{\eta}_k|_j \geq \norm{\bm{\eta}_k}_\infty - 2\Delta/(k + 2) \right\}$
        \State 1.b. Update active indices:
        \State \qquad $\mathcal{S}_{k+1} \gets \mathcal{S}_{k} \cup \mathcal{I}_{k+1}$
        \vspace{6pt}
        \State 2. Update active weights:
        \State \qquad $\mathbf{I}_{k+1} \gets \underset{\mathrm{Supp}(\mathbf{I}) \subset \mathcal{S}_{k+1}}{\arg\min} \frac{1}{2} \norm{ \mathbf{V} - \mathbf{\Phi} \mathbf{I}}_2^2 + \lambda \norm{\mathbf{I}}_1$
        \vspace{6pt}
        \State 3. Prune atoms:
        \State \qquad $\mathcal{S}_{k+1} \gets \mathrm{Supp}(\mathbf{I}_{k+1})$
        \vspace{6pt}
        \State 4. Check convergence:
        \State \qquad \texttt{STOP} if a stopping criterion is verified.
    \EndFor

    \\
    \State \textbf{Output:}
    \State \qquad Postprocess $\mathbf{I}^{(k)}$ (\textit{e.g.} convolution with synthetic beam, add residual image)
\end{algorithmic}
\end{algorithm}

\begin{remark}[Positivity constraint]
    The Polyatomic Frank-Wolfe algorithm can optionally enforce positivity on the solutions, a strategy known to produce more accurate radio interferometric images as discussed in Sect.~\ref{remark:pos}. In this case, the candidate atoms are placed only at the positive maximum values of the residual image and the reweighting subproblem is solved with a positivity constraint as well.
\end{remark}

Standard off-the-shelf implementations of radio interferometry operators perform their computations as if the images were dense. Thus, using them with PFW would require many inefficient data conversions between sparse and dense storage of the iterates. To take full advantage of the sparse iterates of PFW, we need to perform more efficient (de-)gridding operations. Our solution is to use an implementation of the forward operator that conveniently handles non-uniform data and keeps sparse inputs and outputs.

\subsection{(De-)Gridding with \texttt{HVOX}}
\label{sec:nufft}

The degridding (computation of the visibilities) and gridding (image synthesis) operations correspond to applying respectively the forward operator $\mathbf{\Phi}$ and its adjoint $\mathbf{\Phi}^*$, as given in equations \eqref{eq:phi-discrete} and \eqref{eq:adjoint-phi}. A characteristic of the PFW algorithm is that the iterates $\mathbf{I}_k$ are sparse images, and thus stored in a sparse format to enhance scalability. Additionally, the reweighting sub-problem of the second step of PFW enforces a constraint on the support which is straightforward to apply with a sparsity-aware forward operator. For these reasons, PolyCLEAN makes use of the \texttt{HVOX} package \citep{kashani2023} to implement the forward operator. This implementation is specifically designed to efficiently handle sparse skies and sparse Fourier planes. Production gridders used in radio astronomy, such as NIFTY \citep{ye2022}, deal only with dense images and thus are not flexible enough to be combined efficiently with PFW. 

Under the hood, \texttt{HVOX} performs a non-uniform Fourier transform, which amounts to evaluating the following sum
\begin{equation*}
    v_{k} = \sum_{j=1}^{M} w_{j} e^{- \mathrm{j} \langle \mathbf{z}_k, \mathbf{x}_{j} \rangle }
\end{equation*}
for $k \in \{1, \ ...\,, L\}$.
The arguments of the transform are $(\mathbf{x}_{j}, w_j) \in \mathbb{R}^3 \times \mathbb{R}$, a collection of $M$ coordinates-intensity pairs representing the input image, and the $\mathbf{z}_k \in \mathbb{R}^3$, which are the $L$ spatial frequencies to compute.  With an appropriate choice of $\mathbf{x}_{j}, w_j$ and $\mathbf{z}_k$, this sum evaluates the action of $\mathbf{\Phi}$ and the outputs $v_{k}$ are the visibility measurements. The image coordinates $\mathbf{x}_{j}$ can be either non-uniform (support of a sparse image) or uniform (dense image), respectively leading to the so-called \textit{type 3} or \textit{type 2} NUFFTs. \texttt{HVOX} conveniently implements the adjoint operator $\mathbf{\Phi}^*$ automatically.

\subsection{Interpretation of the regularization parameter}
\label{sec:interp}
We conclude this section by proposing a way to set and interpret the regularization parameter $\lambda$ of the LASSO problem \eqref{eq:lasso}.

The LASSO solution depends crucially on the value of the regularization parameter $\lambda$, and setting this value is critical to obtain a good reconstruction. Within the Bayesian interpretation, this parameter sets the ratio between the intensity of the measurement  noise and the prior distribution. A high value leads to sparser solutions. However, the range of values that $\lambda$ can take is arbitrary and varies significantly given the problem dimension. Some strategies exist to automatically infer the value of $\lambda$ from the data, for instance including another Bayesian prior on the hyperparameters with a hierarchical model, but we do not consider these extensions in this work. Instead, we use a systematic method to scale $\lambda$ with respect to the problem size, so that it only needs to be fine-tuned manually to account for the noise level. This strategy is commonly used in signal processing and is presented hereafter.

\paragraph{Autoscaling of the penalty parameter}
There exists a critical value $\lambda_\mathrm{max}$ such that, for any larger values $\lambda \geq \lambda_\mathrm{max}$, the only solution to the LASSO problem is null (see for instance \cite[Proposition~II.1]{Koulouri_Heins_Burger_2021}). This value has the following closed-form expression
\begin{equation}
    \lambda_\mathrm{max} = \norm{\mathbf{\Phi}^* \mathbf{V}}_\infty \geq 0.
    \label{eq:lmax}
\end{equation}
For PolyCLEAN reconstructions, we propose to set the value of the regularization parameter as a fraction of this maximum value
\begin{equation}
    \label{eq:lambda}
    \lambda = \alpha \lambda_\mathrm{max},
\end{equation}
with $0 \leq \alpha \leq 1$. The value of $\alpha$ typically belongs to the range $[0.01, 0.10]$ and needs to be specifically adapted by the user depending on the estimated noise intensity.

\paragraph{Interpretation of $\lambda$}
Interestingly enough, we can interpret $\alpha$ as the ratio of measured information that is not explained by the LASSO and remains in the residual. Indeed, at convergence of the algorithm the dual certificate of the problem is computed from the solution and satisfies the conditions given in Sect. \ref{sec:lassoprops}. In particular, when the solution is non-null, equations \eqref{eq:dual-saturation1} and \eqref{eq:dual-saturation2} lead to $\norm{\bm{\mu}_\lambda}_\infty = 1$. Plugging the expression of $\lambda$ from \eqref{eq:lambda}, we obtain
\begin{equation*}
    \alpha = \frac{\norm{\mathbf{\Phi}^*\left( \mathbf{V} - \mathbf{\Phi I}_\mathrm{sol}\right)}_\infty}{\norm{\mathbf{\Phi}^* \mathbf{V}}_\infty},
\end{equation*}
where $\mathbf{I}_\mathrm{sol}$ is the solution returned by the PFW algorithm. More precisely, $\alpha$ becomes the ratio between the absolute maximum intensity of the residual image and the dirty image. When the algorithm stops, all the pixels in the residual image have an absolute intensity smaller than the $\alpha$-fraction of the initial maximum intensity of the dirty image.

\section{Numerical experiments} \label{sec:exp}

\begin{table*}[ht!]
\centering
\caption{Summary of the datasets used in the different numerical experiments in Section~\ref{sec:exp}.}
\label{tab:datasets}
\begin{tabular}{l@{\hskip .8cm}|@{\hskip .8cm}ccc}
\toprule
Dataset & I & II & III\\
\midrule
Subsection & \ref{sec:profile} & \ref{sec:psobs}\ - \ref{sec:dual} & \ref{sec:diffsim} \\
\midrule
\midrule
Source & Point sources & Boötes field & M31\\
\midrule
Data type & Simulation &Actual observations & Simulation \\
\midrule
Antenna layout & SKA-LOW core & LOFAR core stations & SKA-LOW core\\
\midrule
Number of stations & 94 - 362 & 50 & 251 \\
\midrule
Longest baseline (in wavelength) & 195 - 3241 & 4813 & 605\\
\midrule
Frequency (MHz) & 100 & 145.8  & 100 \\
\midrule
Bandwidth (MHz) & 1  & 0.1  & 1 \\
\midrule
Number of integration times & 7 & 72 & 11\\
\midrule
Time resolution (s) & 4800 & 400 & 2880 \\
\midrule
Total number of baselines ($\times 10^3$) & 31 - 457 & 68.4 & 31.6 \\
\midrule
FOV (deg) & 5.0 & 6.0 & 6.5 \\
\midrule
Image size (pix) & 288$\times$288 - 10.000$\times$10.000 & 3072 $\times$ 3072  & 256 $\times$ 256 \\
\midrule
PSNR (dB) & 20 & -- & 20 \\
\midrule
Nominal resolution (arcsec) & see Table~\ref{tab:resolutions} & 14.3 & 113.6 \\
\midrule
Resolution used (arcsec) & see Table~\ref{tab:resolutions} & 7.0 & 91.4 \\
\bottomrule
\end{tabular}
\end{table*}

In this section, we present four experiments that investigated how PolyCLEAN compares to other imaging methods in terms of capabilities and performance.
\begin{itemize}
    \item Subsection \ref{sec:profile} reports on the scalability of PolyCLEAN. We compared the reconstruction time of different methods on simulated measurements of a sky containing point sources with different pixel resolutions.
    \item In subsection \ref{sec:psobs}, we performed comparative reconstructions of actual radio interferometry observations with CLEAN and PolyCLEAN. We demonstrated that these two methods produced sky maps that are visually similar, with the possibility to perform varying reconstruction depths.
    \item In subsection \ref{sec:dual}, we display the dual certificate obtained at convergence of PolyCLEAN on the same observation set. To the best of our knowledge, this kind of uncertainty image had never been used in a radio interferometric imaging context.
    \item Finally, we evaluate PolyCLEAN for reconstruction of diffuse emissions with a sparse dictionary approach in subsection~\ref{sec:diffsim} using simulated measurements of the Andromeda galaxy (M31).
\end{itemize}

As a common base, we used the \texttt{RASCIL} Python package \citep{rascil} to numerically handle the visibilities, images and interferometer configurations. The code to run PolyCLEAN and to reproduce all the figures is available on a dedicated open access repository (\url{https://github.com/AdriaJ/polyclean}). Notably, we implemented PolyCLEAN using Python and its standard scientific computing libraries. We additionally made use of the \texttt{Pyxu} optimization package for conveniently handling arithmetic of linear operators and the optimization sub-routines \citep{pyxu-framework}. We finally relied on the \texttt{WSCLEAN} software \citep{offringa2014} to run the different  CLEAN-based algorithms, namely Cotton-Schwab for point sources reconstruction and Multi-scale for diffuse emissions, that we both refer to as \textit{CLEAN} regardless of the context.

The characteristics of the datasets for the different experiments are summarized in Table~\ref{tab:datasets}.

In our simulations, we considered omnidirectional antennas and assumed the antenna beam to be constant on the limited field of view. This choice does not impact the scalability and speed of the algorithms.

\subsection{Scalability and speed assessment}
\label{sec:profile}

\begin{table}[ht]
    \caption{Resolution of the recovered images when changing the super resolution factor in Sect.~\ref{sec:profile} compared to the nominal resolution. The resolution is defined as the size of a pixel in arcseconds.}
    \label{tab:resolutions}
    \centering
    \begin{tabular}{m{.8cm}|@{\hskip .2cm}m{1.45cm}m{1.5cm}m{1.45cm}m{1.45cm}}
    \toprule
    {$r_\mathrm{max}$ (km)} & \centering Nominal resolution & \centering Resolution SRF 2 & \centering Resolution SRF 5 & \centering Resolution SRF 10 \arraybackslash \\
    \midrule
    0.3 &    353.4 &  125.0 &   50.0 &    25.0 \\
    0.6 &    180.0 &   62.5 &   25.0 &    12.5 \\
    0.9 &    119.5 &   46.9 &   18.8 &     9.4 \\
    1.2 &    109.4 &   41.7 &   16.7 &     8.3 \\
    1.5 &     82.2 &   31.2 &   12.5 &     6.2 \\
    2.0 &     63.7 &   25.0 &   10.0 &     5.0 \\
    3.0 &     43.4 &   16.7 &    6.7 &     3.3 \\
    6.0 &     21.2 &    9.0 &    3.6 &     1.8 \\
    \bottomrule
    \end{tabular}
\end{table}

\begin{figure}[ht]
    \centering
    \includegraphics[scale=.8, trim=0 0 0 0, clip]{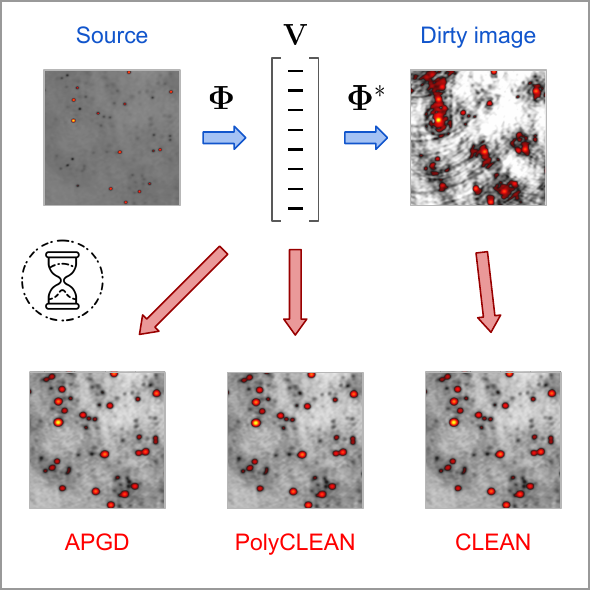}
    \caption{Schematic view of the simulation pipeline for the scalability experiment. The CLEAN procedure starts from the dirty image, while PolyCLEAN and APGD take as input the visibilility measurements $\mathbf{V}$ and instantiate the initial iterate empty.}
    \label{fig:simpipeline}
\end{figure}

\begin{figure}
\centering
\begin{subfigure}{\hsize}
  \centering
  \includegraphics[width=\linewidth, trim=30 30 30 20, clip]{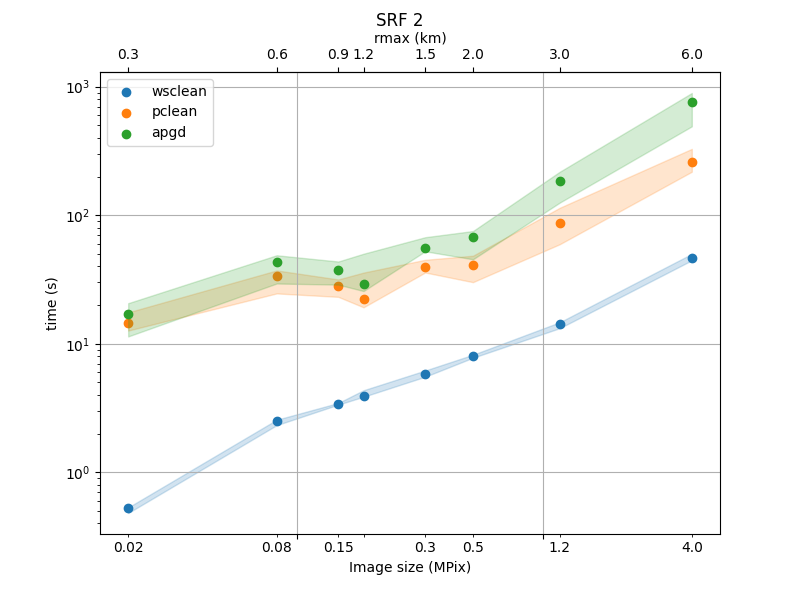}
  \caption{SRF 2}
  \label{fig:srf2}
\end{subfigure}

\begin{subfigure}{\hsize}
  \centering
  \includegraphics[width=\linewidth, trim=30 20 30 20, clip]{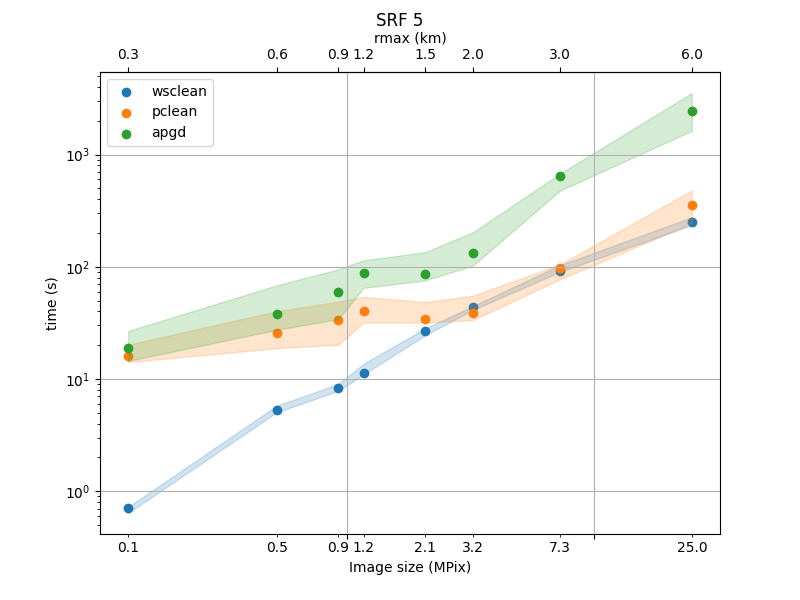}
  \caption{SRF 5}
  \label{fig:srf5}
\end{subfigure}


\begin{subfigure}{\hsize}
  \centering
  \includegraphics[width=\linewidth, trim=30 20 30 20, clip]{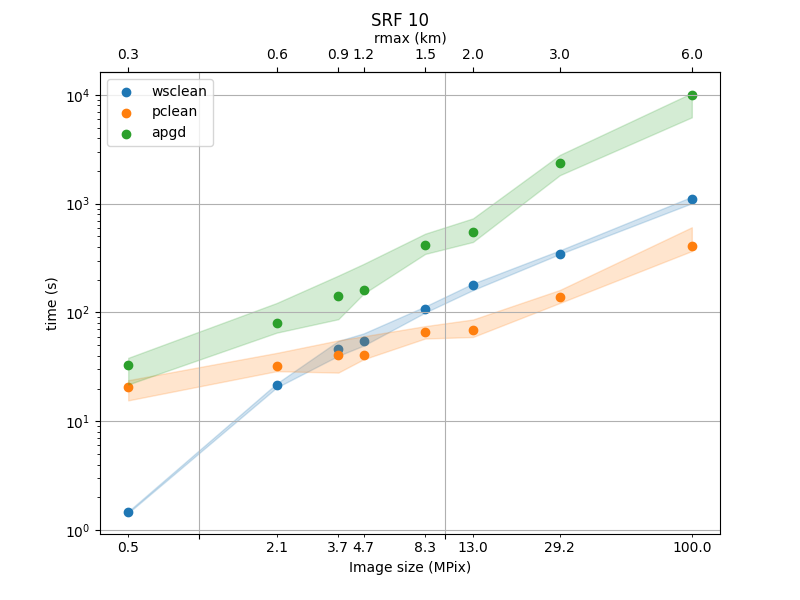}
  \caption{SRF 10}
  \label{fig:srf10}
\end{subfigure}

\caption{Reconstruction time for different super resolution factors (SRF). PolyCLEAN consistently outperforms the dense APGD solver. CLEAN is faster than PolyCLEAN for low SRF factor but gets outperformed when the pixel size decreases. The PolyCLEAN curve, seen as reconstruction time with respect to maximum baseline, is almost identical on the three plots irrespective of the SRF factor.}
\label{fig:scaling}
\end{figure}

We investigated the scalability of PolyCLEAN by performing image reconstructions over simulated measurement sets of increasing size, and compared the solving time of PolyCLEAN against two other methods: CLEAN and APGD \citep{liang2022}. As already described, CLEAN produces a matching pursuit solution to the inverse problem, demonstrating good performance in terms of speed and result quality, thanks to years of continuous improvement. The other contender, APGD, is a proximal optimization algorithm. We used it to solve the same LASSO problem as PolyCLEAN so that the imaging inverse problem was regularized in the same manner using different numerical solvers. Notably, we recall that APGD does not use sparse iterates.

\paragraph{Simulation of the problem}
A given area of the sky was simulated and we performed many visibility measurements of this sky using interferometers of increasing station count. The reconstructed images always covered the same area of the sky, but the resolution increased with the size of the telescope array. Longer baselines means more visibility measurements and more pixels in the reconstruction, hence the inverse problem considered gets larger ($L$ and $N$ increase in $\mathbf{\Phi}: \mathbb{R}^N \to \mathbb{C}^L$). The simulation parameters are summarized in the first column of Table~\ref{tab:datasets}. The exact problem dimensions across the experiment are reported in the table in Appendix~\ref{app:pbm-size}.

The visibilities were corrupted with additive Gaussian noise with PSNR nominatively set to $20$ dB.  The field of view had a fixed angular size of $5^\circ\times 5^\circ$. $200$ point sources were randomly placed on this sky. The source intensities were drawn with a log-normal distribution. The baselines reproduced a configuration of the core SKA-LOW interferometer. Only stations closer than a given radius $r_\mathrm{max}$ from the center of the interferometer were considered. A low value of $r_\mathrm{max}$ only samples the lowest spatial frequencies while higher frequencies are observed as the radius increases.

\paragraph{Resolution of the images}
Table~\ref{tab:resolutions} reports on the pixel size of the experiments. For a given field of view, the user-defined image resolution plays a significant role in solver scalability. A commonly chosen pixel size is the so-called \textit{nominal resolution} $\delta \ell_0 = 1/(3B_0)$ (in radians) where $B_0 = \max_{k=\{1, \dots, L\}} \norm{(u_k, v_k, w_k)}_2$ is the largest baseline. However, it has been shown that nonlinear reconstruction techniques that rely on sparse priors are able to resolve objects at finer resolutions \citep{honma2014, Akiyama_2017, dabbech2018, Pan_Simeoni_Hurley_Blu_Vetterli_2017}. We thus chose to perform the experiments at different fine resolutions which correspond to approximately $2, 5$ and $10$ times smaller pixel size than $\delta \ell_0$. We refer to these setups as SRF2, SRF5 and SRF10 (for \textit{Super-Resolution Factor}).

\paragraph{Solvers parameters}
For a fair comparison between the different methods, the algorithms were configured so that they output reconstructions with similar quality by adjusting the reconstruction parameters as well as the stopping criteria. The reasons for those choices are as follows.

Regarding quality assessment, model images returned by different solvers are difficult to compare to the simulated ground truth due to their pixel-sparse nature. Any localization error excessively impacts regular image metrics such as \textit{mean square error} (MSE) or \textit{mean absolute deviation} (MAD). Indeed, even a one pixel error on the reconstruction leads to high metrics values although the recovered intensity may be correct.  For this reason, we first smoothed the output models by convolving them with the \textit{CLEAN beam} -- a Gaussian kernel fitted on the main lobe of the PSF -- before computing MSE and MAD metrics. The resulting metrics are then affected by localization and intensity errors in a more balanced manner. These were then used to ensure that CLEAN and PolyCLEAN produced similar quality images. The average values of these metrics along repetitions of this experiment are indeed comparable and provided in Appendix \ref{app:metrics}. \

The choice of an appropriate stopping criterion is also critical to obtain relevant time comparisons. In constrast to CLEAN, reconstructions obtained with LASSO solvers usually benefit from finer convergence, at the cost of longer solving time. A compromise is needed between time efficiency and accuracy. The following are the three algorithms and their respective parameters:
\begin{itemize}
    \item Cotton-Schwab CLEAN was performed with minor cycles using \texttt{WSCLEAN} with a \texttt{mgain} parameter of $0.7$ and a cycle \texttt{gain} parameter of $0.1$. The \texttt{niter} parameter was set high and never reached. \texttt{WSCLEAN} stopped when the residual noise was smaller than some threshold. As recommended by the authors \citep{offringa2014}, we used the \texttt{auto-threshold} argument that self-determined the value of the threshold with respect to the standard deviation of the image. In this experiment, the \texttt{auto-threshold} was set to $1$.
    \item PolyCLEAN solves a LASSO problem, hence the quality of the reconstruction and the residual noise directly depend on the value of the regularization parameter. We used the rule proposed in Sect.~\ref{sec:interp} and set $\lambda = \alpha \lambda_\mathrm{max}$ with $\alpha = 0.01$. This value was handpicked to provide similar reconstruction quality between CLEAN and PolyCLEAN. The algorithm stopped when the relative improvement of the LASSO objective function got smaller than $10^{-4}$.
    \item APGD solved the same LASSO problem as PolyCLEAN, so the regularization parameter had the same value. The algorithm stopped as soon as the value of the objective function reached the minimum obtained with PolyCLEAN so that both solvers had equal treatment with respect to the objective function. APGD also used the \texttt{HVOX} implementation of the forward operator.
\end{itemize}

\paragraph{Implementation}
CLEAN and PolyCLEAN are at different levels of maturity. The implementation of \texttt{WSCLEAN} is state-of-the-art C++ and has benefited from continuous development. In contrast, PolyCLEAN and APGD are based on the Python package \texttt{Pyxu}. Only the computation of the forward operator has been optimized using \texttt{HVOX}: the rest of the numerical treatment was done with \texttt{numpy}. Thus, the good performance demonstrated by PolyCLEAN in these experiment augured well for the use of sparsity-based algorithms to improve the scalability of optimization-based methods.

\paragraph{Simulation results}
The scaling behavior of the three algorithms is shown in Fig.~\ref{fig:scaling}, represented by the slope of the curves. The upper x-axis represents the radius of the interferometer, while the lower x-axis displays the corresponding image size (in megapixels). Both are represented on a logarithmic scale. The reconstruction times are provided on the y-axis on a logarithmic scale as well. The markers represent the median timings over $10$ repetitions with the same interferometer radius, while the shaded areas represent the inter-quartile range and illustrate the variability over repetitions. The reconstructions were run on a workstation containing a processor Intel Core i9-10900X 10-core CPU and 128 GB of DDR4 memory. Some reconstruction examples with PolyCLEAN for various problem sizes are illustrated in Appendix~\ref{app:reco}.

The experiment demonstrates that PolyCLEAN performed significantly better than APGD for all problem dimensions considered. This result strengthens the observations made in the original PFW article \citep{Jarret_Fageot_Simeoni_2022}.

Compared to CLEAN, PolyCLEAN also demonstrated good scalability in the high dimensional contexts considered. Indeed, we can see on the graphs \ref{fig:srf5} and \ref{fig:srf10} that PolyCLEAN matched and even outperformed \texttt{WSCLEAN} for images of size $3$ megapixels and more. However, CLEAN was consistently faster than the LASSO-based methods for small size images. For the low super-resolution factor considered in graph \ref{fig:srf2}, CLEAN also significantly outperformed the optimization methods.

One possible explanation for the three plots is the sparsity of the simulated source images. Indeed, there were only $200$ point sources spanning $200$ pixels, independent of image sizes that varied from a few thousands pixels up to $100$ megapixels. Therefore the source images got sparser as the interferometer size increased. In this context, the sparse processing and sparsity-aware operators of PolyCLEAN enabled efficient high resolution imaging to be performed in observational setups with large baselines, increasingly outperforming the dense operations of CLEAN.

Interestingly, we notice that the speed curves of PolyCLEAN are nearly identical on all subplots \ref{fig:srf2}, \ref{fig:srf5} and \ref{fig:srf10} irrespective of the chosen SRF. This observation shows that the reconstruction time of PolyCLEAN merely depends on the intrinsic resolution of the solution (correlated to the largest baseline) and was little affected by the actual reconstruction resolution.

\subsection{{LOFAR observations of the Bo\"otes field}}
\label{sec:psobs}

\begin{figure}
\centering
  \includegraphics[width=\hsize, trim=0 0 0 26, clip]{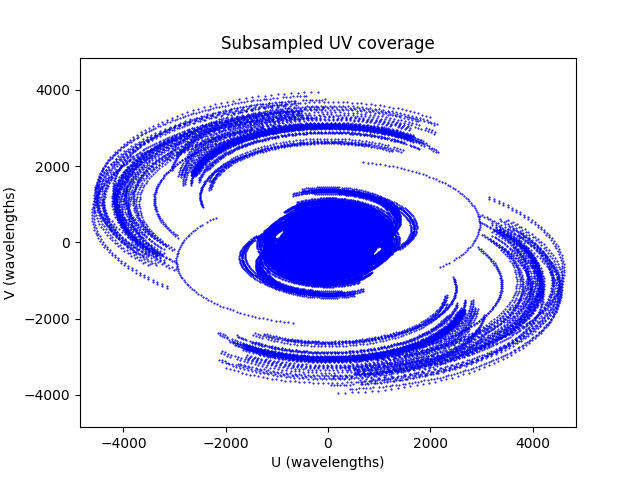}
  \caption{UV coverage of dataset II, coordinates are given in multiple of observation wavelength.}
  \label{fig:uvcov}
\end{figure}

\begin{figure}
\centering
  \includegraphics[width=\hsize, trim=60 0 0 40, clip]{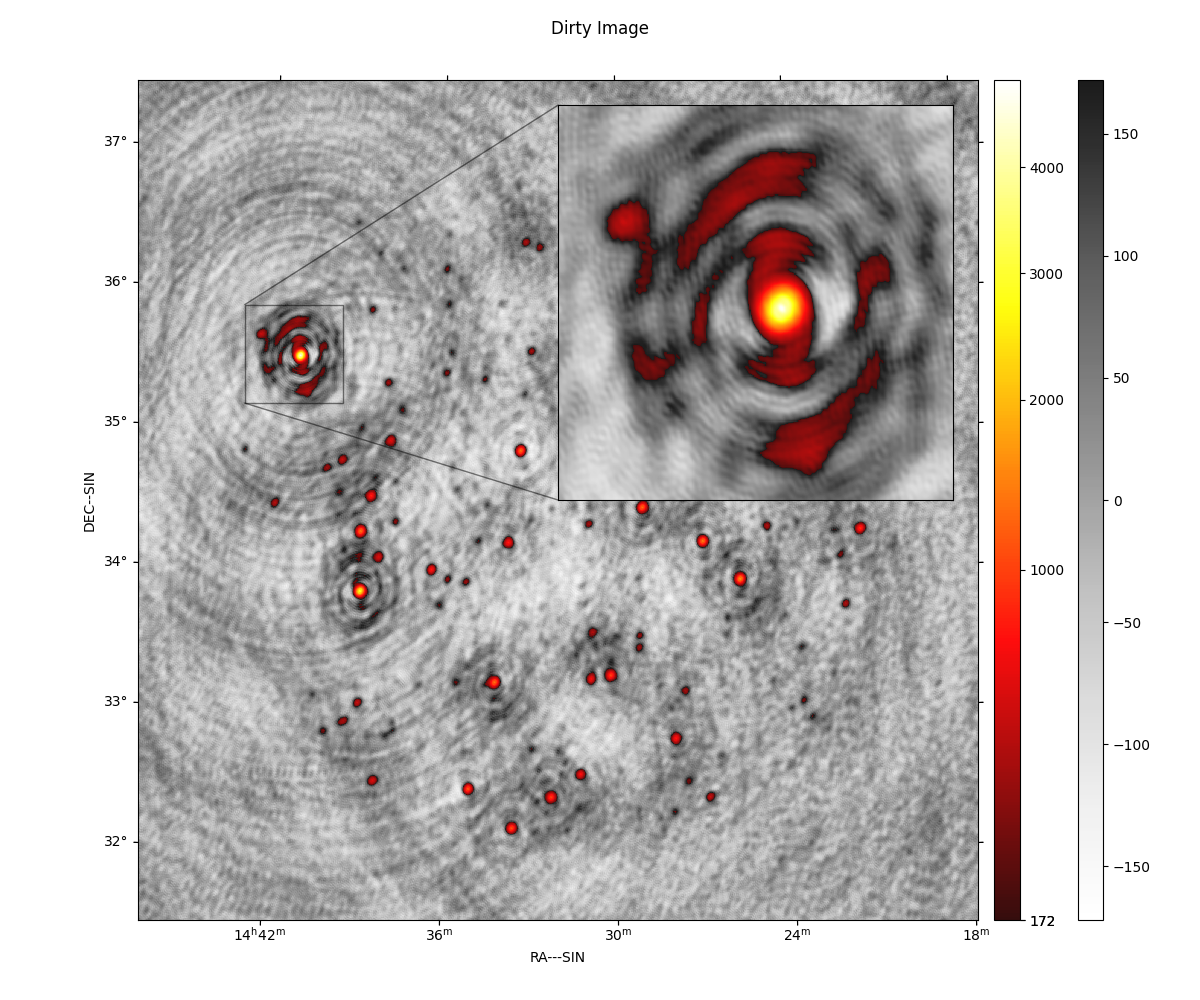}
  \caption{Dirty image of the Bo\"otes field obtained with dataset II. Reconstructions are provided in Figs.~\ref{fig:lofar-comp} and \ref{fig:lofar-res}.}
  \label{fig:dirty}
\end{figure}

\begin{figure}
\centering
  \includegraphics[width=\hsize, trim=60 0 0 40, clip]{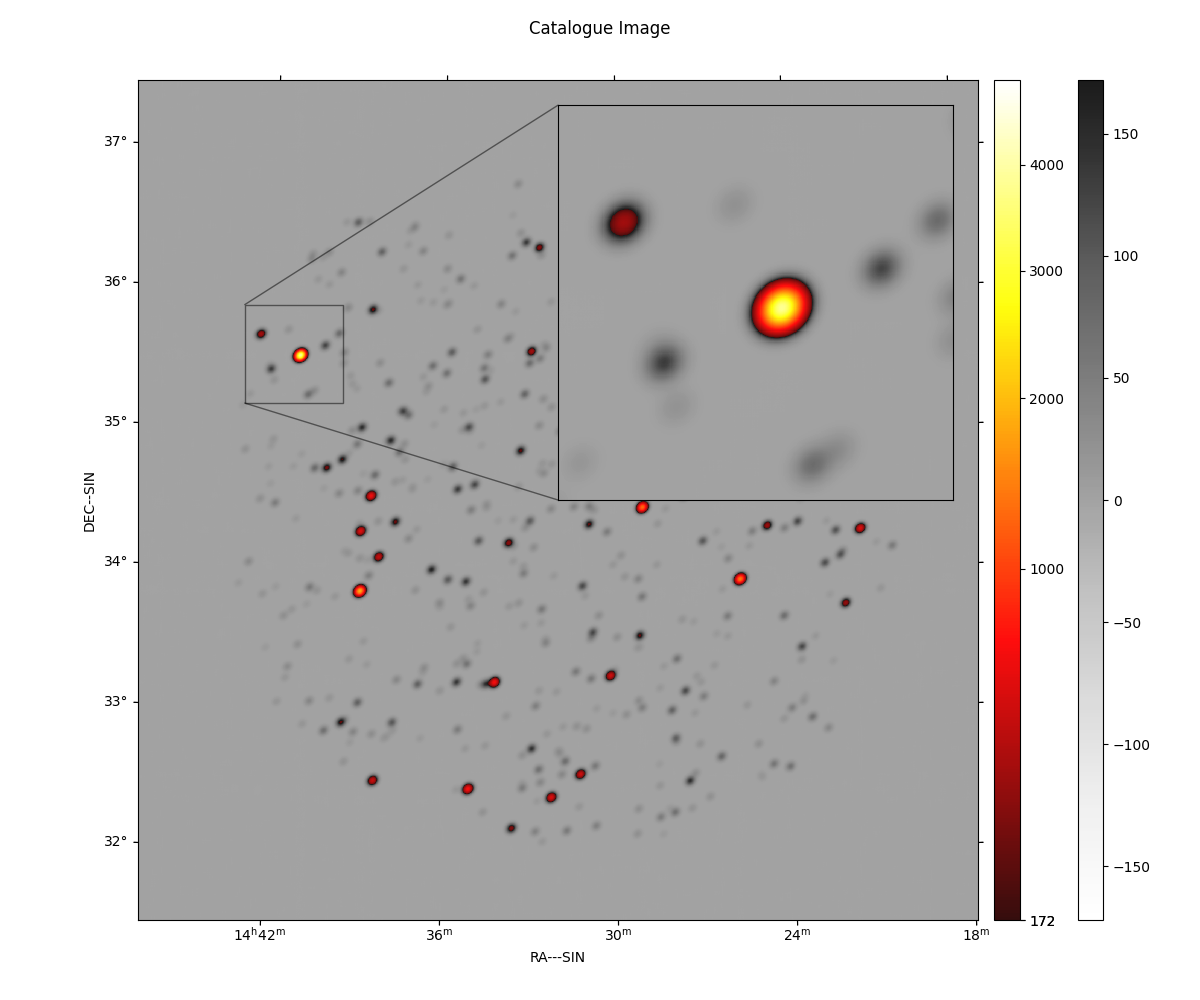}
  \caption{Catalog image of the Bo\"otes field corresponding to dataset II, constructed using the CLEAN beam computed with the stations of dataset II.}
  \label{fig:catalog}
\end{figure}

\begin{figure*}
    \centering
    \resizebox{0.92\hsize}{!}{
    \includegraphics[trim=20 18 20 5, clip]{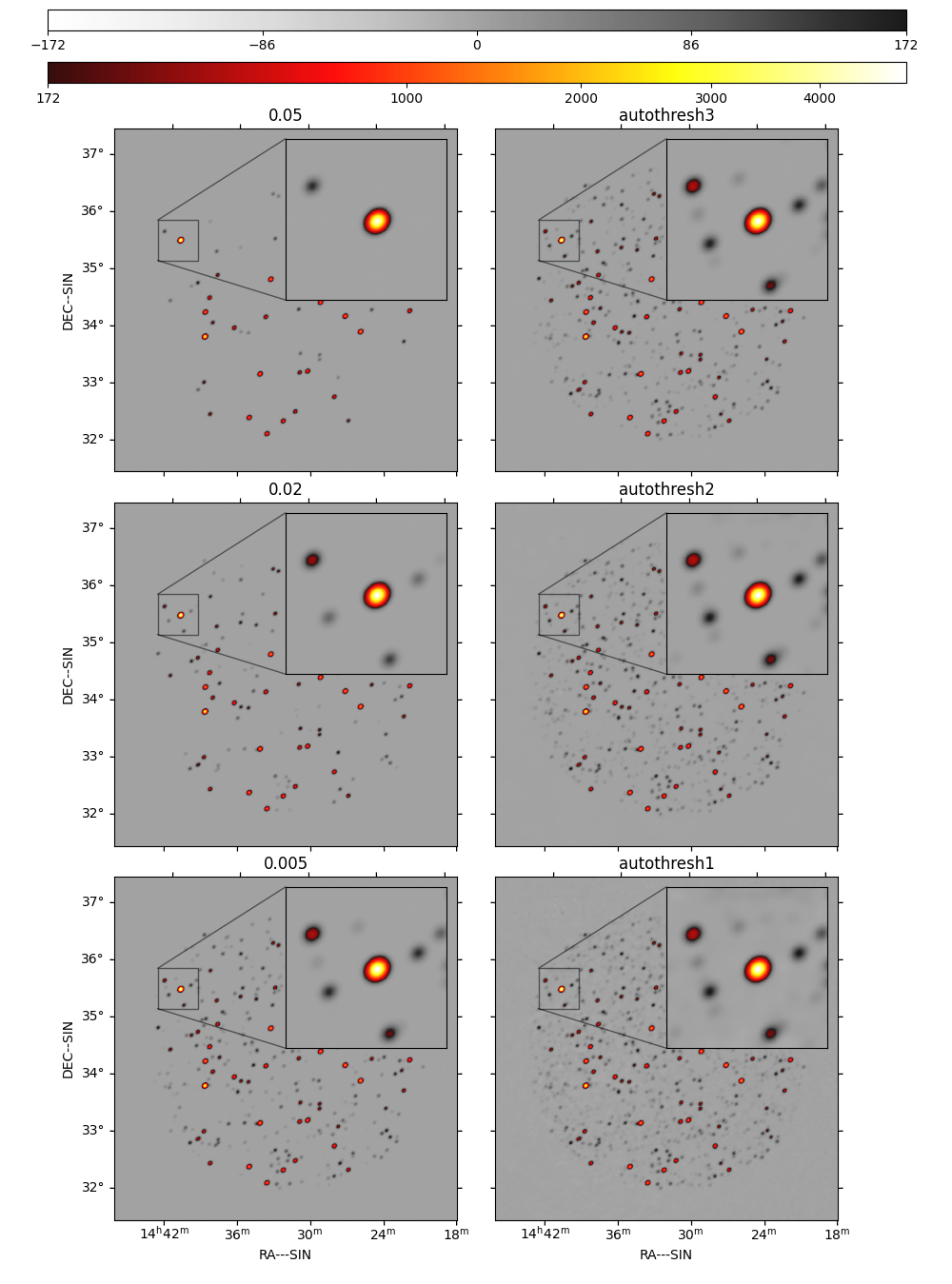}}
    \caption{Comparison without residual. \textit{Left:} PolyCLEAN images, with decreasing $\alpha$ values that determine the level of regularization. \textit{Right:} \texttt{WSCLEAN} images with decreasing level of residual noise at stop, determined by the \texttt{auto-threshold} parameter. As mentioned, images should not be directly compared side-by-side as they do not present the same level of regularization.}
    \label{fig:lofar-comp}
\end{figure*}

\begin{figure*}
    \centering
    \resizebox{0.92\hsize}{!}{
    \includegraphics[trim=20 18 20 5, clip]{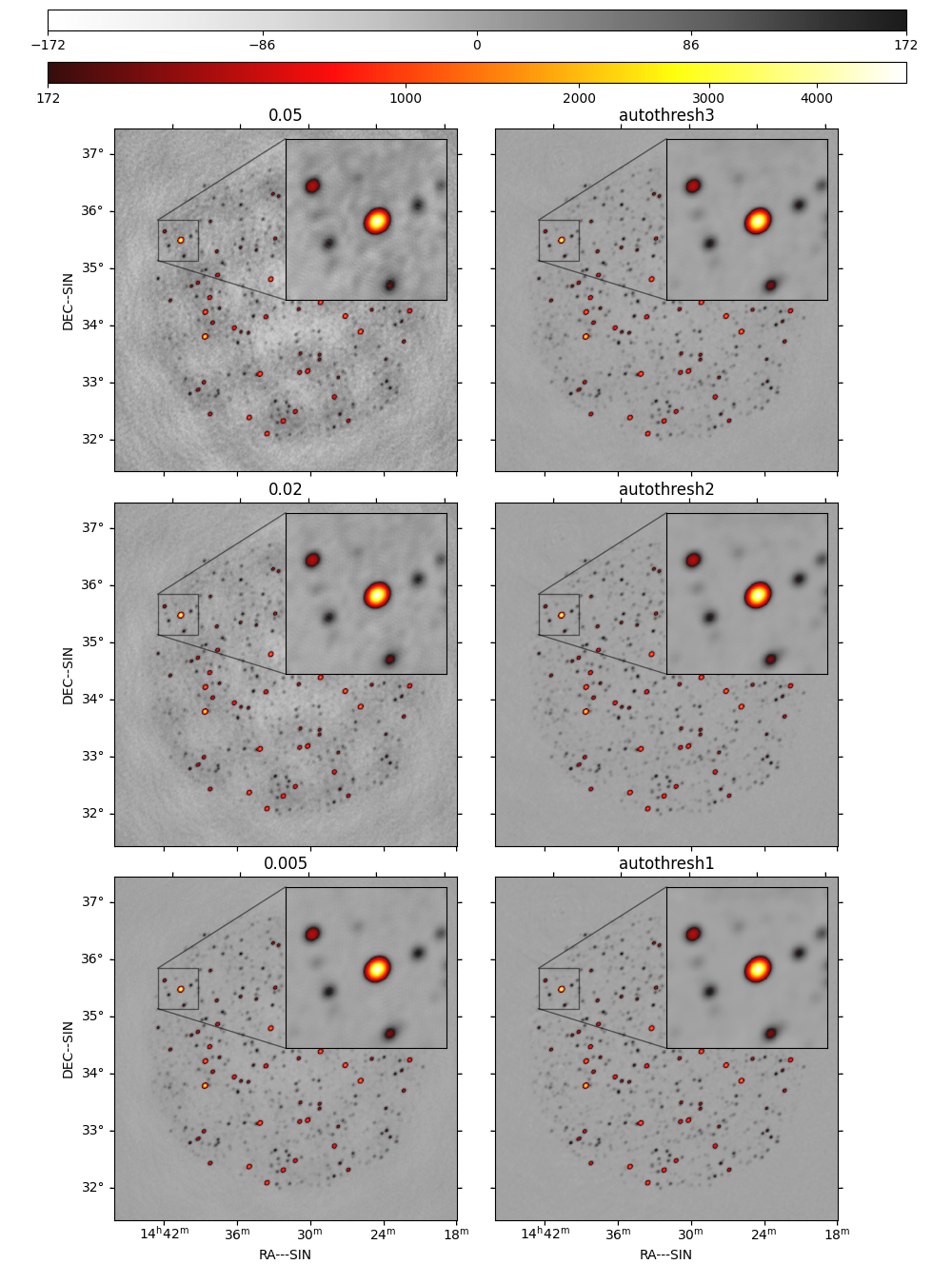}}
    \caption{Comparison with dirty residual. Same experiment as Fig.~\ref{fig:lofar-comp}. PolyCLEAN images are on the left while \texttt{WSCLEAN} ones are on the right.}
    \label{fig:lofar-res}
\end{figure*}

The second experiment compared the images reconstructed by PolyCLEAN and CLEAN when applied to the same dataset of the Boötes field observations. Consistent with experiments using other sparse optimization techniques -- \cite{Wiaux_Jacques_Puy_Scaife_Vandergheynst_2009, Li_Cornwell_Hoog_2011, Carrillo_McEwen_Wiaux_2012} to name a few, PolyCLEAN matched the image quality obtained by CLEAN. Additionally, we illustrate the ability of PolyCLEAN to perform image reconstruction at different levels of residual noise, similar to CLEAN.

\paragraph{The observations}
Dataset II is presented in Table \ref{tab:datasets}. It contains actual observations made with LOFAR over an observation period of 8 hours with an integration time of $8$ seconds. The visibilities were subsampled in time so that only 2\% of the dataset was used for image reconstruction, with one measurement every 400 seconds (one measurement kept every $50$ integration times). The measurements were obtained with LOFAR using the 48 core stations and the two closest ones out of the core.  The resulting UV coverage is displayed in Fig.~\ref{fig:uvcov}.  The field of view was a square covering 6°$\times$6° of the sky, and was imaged over $3072 \times 3072$ pixels, giving a pixel resolution of $7.0$ arcseconds. Comparatively, the nominal resolution for the considered observations was $14.3$ arcseconds (which is a super resolution factor of 2.03). The dirty image resulting from this measurement set is provided in Fig.~\ref{fig:dirty}. For visualization purposes, two scales were used. The background components are displayed in a symmetric linear white-to-black colormap, centered around $0$, while the more intense foreground components are represented with a \textit{hot} colormap (black - red - yellow) with a square root scale. The square root law conveniently compresses the higher end of the dynamic range.

\paragraph{Reconstruction parameters}
The following configuration was used:
\begin{itemize}
    \item PolyCLEAN was used in the same setup as in the first experiment: \texttt{HVOX} was employed to perform the (de-)gridding operations and the regularization parameter $\lambda$ was set to $\alpha \lambda_\mathrm{max}$ for some $0 < \alpha < 1$. The algorithm again stopped when the relative improvement on the objective function became smaller than $10^{-4}$. 
    \item CLEAN was also applied using the standard Cotton-Schwab version implemented by \texttt{WSCLEAN}. The \texttt{auto-threshold} parameter determined the stopping criterion, as recommended by the authors.
\end{itemize}
In this experiment, the $\alpha$ regularization parameter for PolyCLEAN took the values $[0.05, 0.02, 0.005]$ while the \texttt{autothreshold} for CLEAN explored the range $[3\sigma, 2\sigma, 1\sigma]$ (with $\sigma$ the standard deviation of the dirty image). The values of $\alpha$ were handpicked so that PolyCLEAN visually matched the reconstruction depth of CLEAN in each case.

\paragraph{Image reconstruction}
Both CLEAN and PolyCLEAN produced a \textit{model image} of point sources which can be very sparse and difficult to interpret. We used the standard radio interferometry pipeline to produce the images provided in Figs.~\ref{fig:lofar-comp} and \ref{fig:lofar-res}. The same colormap as the dirty image is used for display. First, a \textit{CLEAN beam} was computed by fitting a Gaussian kernel to the main lobe of the PSF. The model images were then convolved with this CLEAN beam, leading to Fig.~\ref{fig:lofar-comp}. The dirty residual was added in Fig.~\ref{fig:lofar-res} as it may contain some remaining information. Note that the dirty residual also contained most of the noise isolated during reconstruction, hence it needs to be interpreted with care.

\begin{table}[t]
    \centering
    \begin{subtable}{\linewidth}
    \centering

    \caption{MSE scores}

    \begin{tabular}{m{1.5cm}|m{1.2cm}||m{1.2cm}|m{1.5cm}}
        \toprule
         \multicolumn{2}{c}{PolyCLEAN} & \multicolumn{2}{c}{\texttt{WSCLEAN}} \\
         \centering $\alpha$ & \centering MSE & \centering MSE & \centering autothresh \arraybackslash \\
         \midrule
         \raggedright \textbf{0.05}  & \centering \textbf{342.2} & \centering 617.8 & \raggedleft 3 \arraybackslash \\
         \raggedright 0.02  & \centering 478.8 & \centering 651.6 & \raggedleft 2 \arraybackslash \\
         \raggedright 0.005 & \centering 613.4 & \centering 671.5 & \raggedleft 1 \arraybackslash \\
         \bottomrule
    \end{tabular}
    \end{subtable}\par

    \begin{subtable}{\linewidth}
    \caption{MAD scores}
    \centering
    
    \begin{tabular}{m{1.5cm}|m{1.2cm}||m{1.2cm}|m{1.5cm}}
        \toprule
         \multicolumn{2}{c}{PolyCLEAN} & \multicolumn{2}{c}{\texttt{WSCLEAN}} \\
         \centering $\alpha$ & \centering MAD & \centering MAD & \centering autothresh \arraybackslash \\
         \midrule
         \raggedright \textbf{0.05}  & \centering \textbf{2.32} & \centering 2.91 & \raggedleft 3 \arraybackslash \\
         \raggedright 0.02  & \centering 2.37 & \centering 3.46 & \raggedleft 2 \arraybackslash \\
         \raggedright 0.005 & \centering 2.74 & \centering 4.70 & \raggedleft 1 \arraybackslash \\
         \bottomrule
    \end{tabular}
    \end{subtable}

    \caption{MSE and MAD between the reconstructions and a catalog image synthesized with the same CLEAN beam. The reconstructions considered are the ones provided in Fig.~\ref{fig:lofar-comp} with the same reconstruction parameters (respectively $\alpha$ and \texttt{autothresh}).}
    \label{tab:ps:metrics}
\end{table}

\paragraph{Analysis}
At first glance the reconstructions of PolyCLEAN and CLEAN look extremely similar. Both methods are able to perform coarse and fine reconstructions by adjusting the parameters to control the level of residual noise. Indeed, the first row of Fig.~\ref{fig:lofar-res} contains the most residual noise which leads to a grainy gray background. This background becomes smoother on the lower rows as the residual is reduced in intensity. 

LASSO solutions are known to introduce a bias in recovered intensities such that some sources may undergo a shrinkage phenomenon which strengthens with the value of the regularization parameter. Additionally, CLEAN may introduce many spurious sources when trying to completely annihilate the residual. Both effects were however smoothed out when convolving with the CLEAN beam, as the latter has large spatial extent compared to the spread of the observed stars.

Note that the reconstructed images are not directly comparable side-by-side as the regularization parameters chosen for PolyCLEAN do not relate to the stopping criteria of \texttt{WSCLEAN}. Hence, each of the six displayed images has a different level of residual image.

We evaluated the quality of the reconstructions by comparing them to the catalog provided by \cite{williams2016}. To do so, we synthesized a \textit{reference} image by convolving the sources from the catalog with the CLEAN beam from the observations of dataset II, resulting in the image provided in Fig.~\ref{fig:catalog}. The MSE and MAD between the reconstructions and this synthetic reference image are reported in Table~\ref{tab:ps:metrics}. Both metrics indicate that the PolyCLEAN images are more faithful to the reference than the images obtained with PolyCLEAN, consistently reaching the best performance for $\alpha = 0.05$. For completeness, the difference images between the reconstructions and the reference are provided in Appendix~\ref{app:diff-obs}.

\subsection{Dual certificate}
\label{sec:dual}

\begin{figure*}
    \sidecaption
    \includegraphics[width=12cm, trim=100 40 10 70, clip]{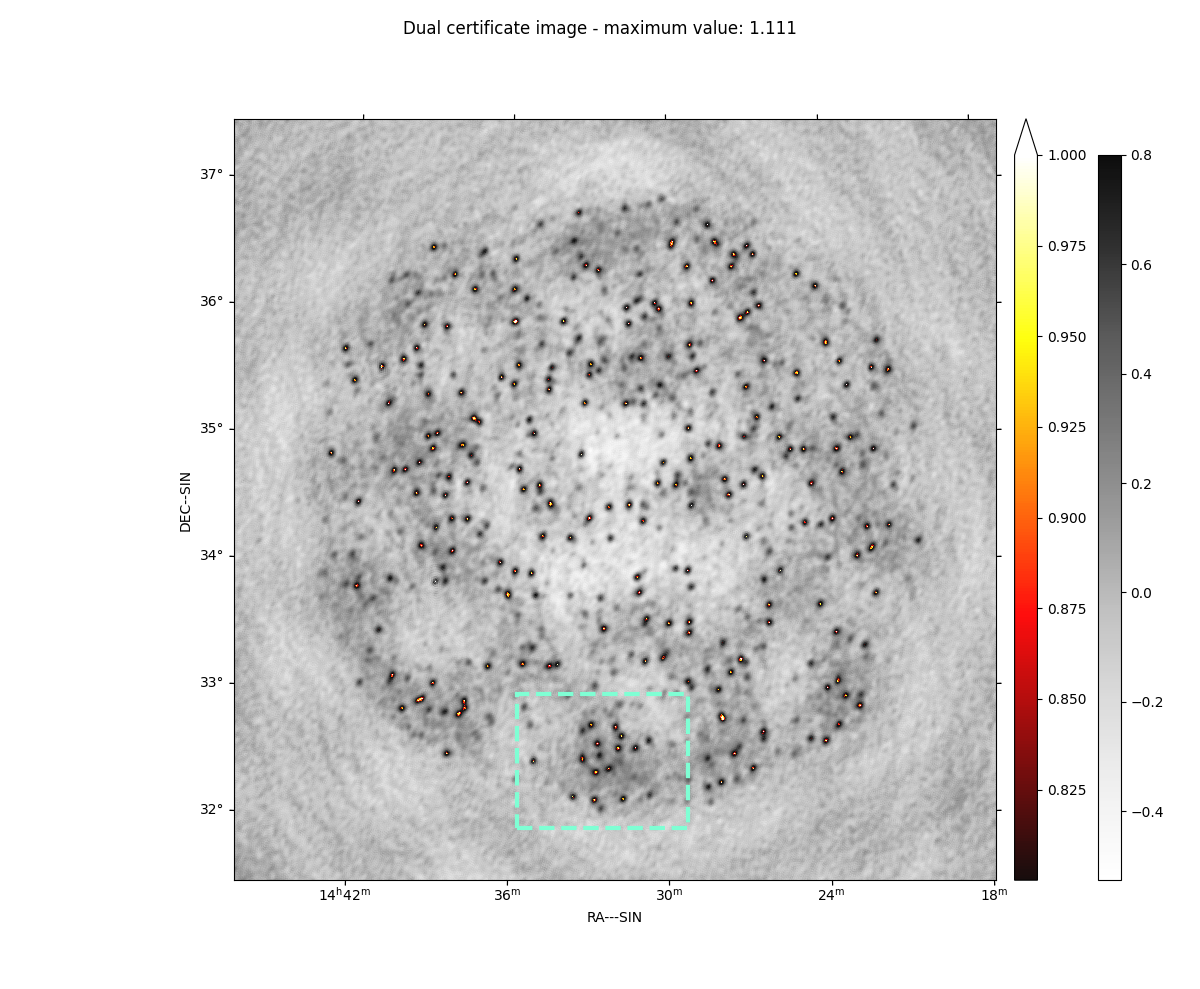}
    \caption{Dual certificate $\bm{\mu}_\lambda$ obtained on the Bo\"otes field reconstruction with parameter $\alpha=0.02$. Figure~\ref{fig:zoom-certificate} shows a zoom-in on the boxed region. This certificate corresponds to the reconstructions displayed on left side, second row in Figs.~\ref{fig:lofar-comp} and~\ref{fig:lofar-res}.}
    \label{fig:dual_cert}
\end{figure*}

\begin{figure*}
\makebox[\textwidth][c]{
\begin{subfigure}{.5\hsize}
  \centering
  \includegraphics[width=\hsize, trim=0 0 0 50, clip]{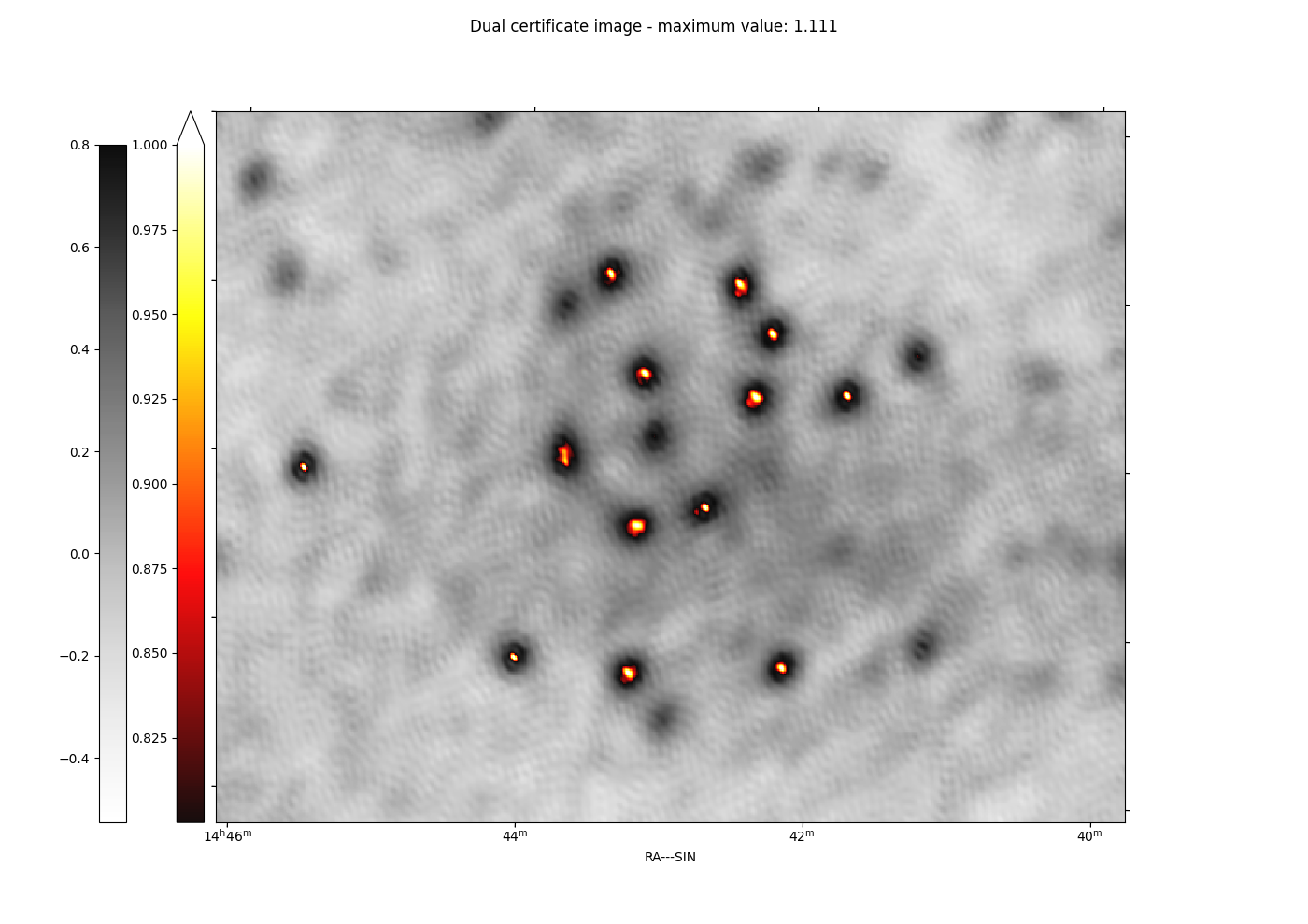}
  \caption{Dual certificate}
  \label{fig:zoom:dc}
\end{subfigure}
\hfill
\begin{subfigure}{.5\hsize}
  \centering
  \includegraphics[width=\hsize, trim=0 0 0 50, clip]{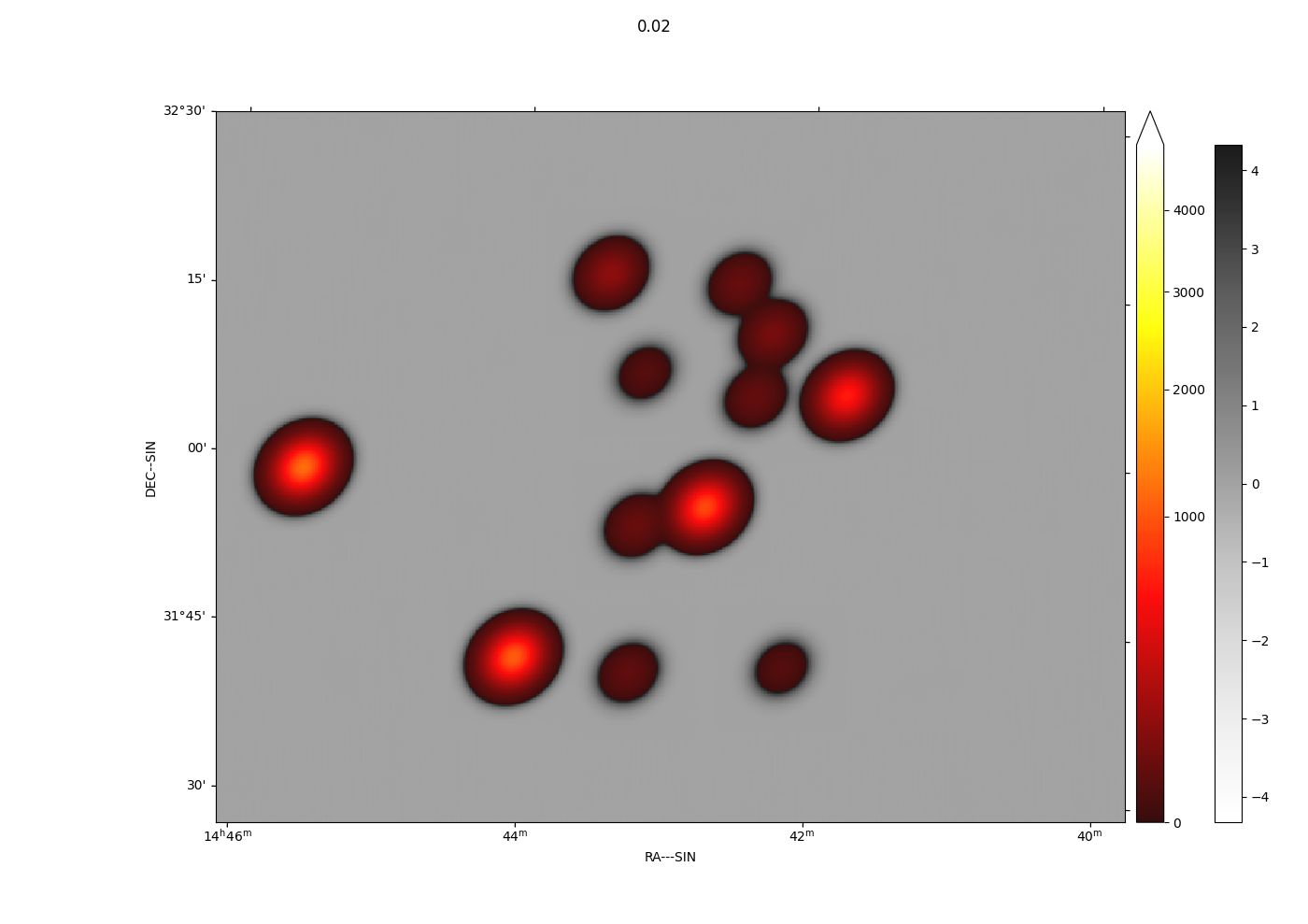}
  \caption{PolyCLEAN}
  \label{fig:zoom:pclean}
\end{subfigure}
}
\caption{Zoom-in: comparison of the dual certificate (a) and the PolyCLEAN reconstruction (b) with the same value of $\alpha$. The reconstruction is convolved with the CLEAN beam.}
\label{fig:zoom-certificate}
\end{figure*}

As described in Sect.~\ref{sec:lassoprops}, PolyCLEAN solves a LASSO problem for which we can compute the dual certificate at convergence $\bm{\mu}_\lambda$ with equation \eqref{eq:dual_certif}. $\bm{\mu}_\lambda$ is an image with the same dimension as the sky image with values in the range $[-1, 1]$\footnote{When a positivity constraint is enforced, the dual certificate may have values smaller than $-1$, but still saturates at maximum $1$.}.

For imaging purposes, we are interested in the saturation set $\mathcal{D}_\lambda$ (defined in equation~\eqref{eq:saturationset}) as this set is made of all the candidate reconstructed source locations. Remarkably, when the LASSO problem admits more than one solution, the dual certificate does not depend on the actual reconstruction returned by PolyCLEAN and delimits every possible reconstruction this LASSO could have produced.

Figure~\ref{fig:dual_cert} shows the dual certificate associated with the LASSO reconstruction of the Boötes field from the previous section (dataset II), with a penalty parameter of $\lambda = 0.02 \lambda_\mathrm{max}$. Two colorbars are given, both on a linear scale. The warm colorbar represents the highest values of the dual certificate, above a threshold at $0.8$. The bright areas can be used as a proxy to identify the saturation set of the dual certificate. Indeed, the convergence is subject to numerical approximations so that the dual certificate does not reach the exact saturation set of the problem.

 We observe that the bright areas indeed correspond to the center of the sources reconstructed in the previous section (see Fig. \ref{fig:lofar-res}). An interesting phenomenon however is that the size of these bright blobs is smaller than the areas of the sources recovered after convolution with the CLEAN beam. In this sense, the dual certificate is more resolved around the recovered source locations, and the area of each blob can be interpreted as an uncertainty estimate on the position. When these uncertainty areas are small, the reconstructed sources are precisely identified by the LASSO, with more confidence than what we obtain in the CLEAN-like reconstructed image after convolution with the CLEAN beam.

Figure~\ref{fig:zoom-certificate} provides a zoom on a small set of point sources. The left panel contains the dual certificate, while the right panel illustrates the PolyCLEAN reconstruction. We can see that the dual certificate identifies more sources than the reconstruction. However, it does not inform as to the intensity of these sources, which needs to be read from the reconstructed maps. Strong sources appear brighter and more resolved on the dual certificate than fainter ones as they are easier to recover and localize.

\begin{figure}
    \centering
    \includegraphics[width=\hsize, trim=20 30 20 50, clip]{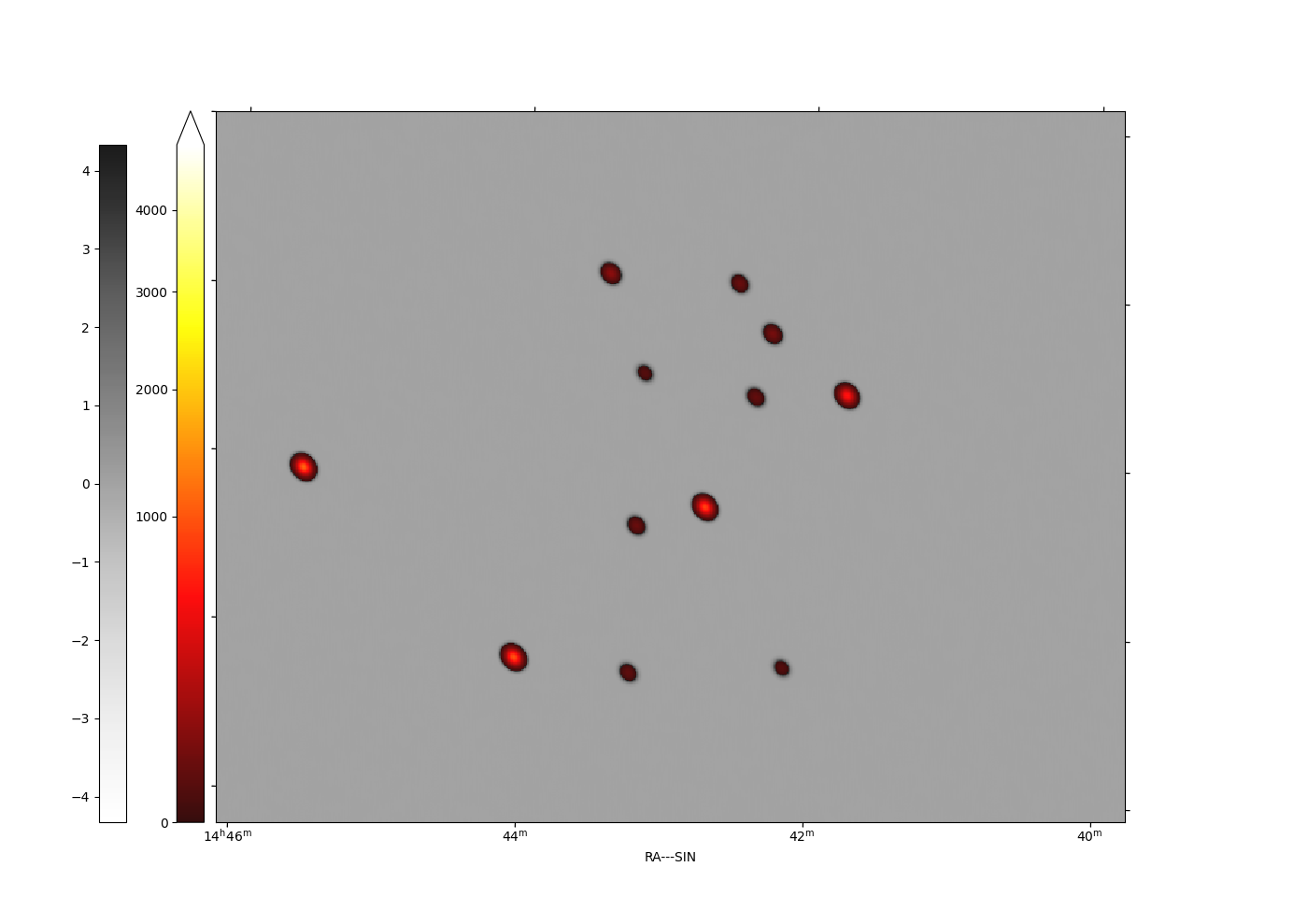}
    \caption{PolyCLEAN reconstruction convolved with a representation beam computed from the dual certificate. The area represented is the same as in Fig.~\ref{fig:zoom-certificate} using the same colormap as Fig.\ref{fig:zoom:pclean}. The threshold to the dual certificate was set to $0.9$.}
    \label{fig:certif-beam}
\end{figure}

Many use cases for the dual certificate can be considered. For instance, computing the area of each bright zone can provide a way to quantify the uncertainty on the source location. For illustration, we propose a procedure to build a representation beam, similar to the CLEAN beam, that encompasses the actual critical resolution of the LASSO reconstruction. The steps are as follows:
\begin{enumerate}
    \item Threshold the dual certificate to a target level of accuracy, \textit{e.g.} set to zero the pixels with value lower than $0.9$;
    \item Compute the \textit{Power Spectral Density} by taking the square of the magnitude of the \textit{Discrete Fourier Transform}, which corresponds to the Fourier transform of the autocorrelation of the dual certificate;
    \item Perform an \textit{Inverse Discrete Fourier Transform}. We obtain a sort of PSF whose main lobe has a spatial extension induced by the size of the intensity peaks of the dual certificate;
    \item Fit a Gaussian kernel to the main lobe and normalize to $1$ similarly to the computation of the CLEAN beam.
\end{enumerate}
With this procedure, the spread of the representation beam is faithful to the resolution of the LASSO solution at convergence of the minimization algorithm. Figure~\ref{fig:certif-beam} displays the PolyCLEAN reconstruction once convolved with the representation beam obtained with this procedure. The image is significantly more resolved than with the CLEAN beam while preserving the intensity values.

\subsection{Diffuse emissions in simulations}
\label{sec:diffsim}

\begin{figure}[t]
    \centering
    \includegraphics[width=\hsize, trim=20 20 20 50, clip]{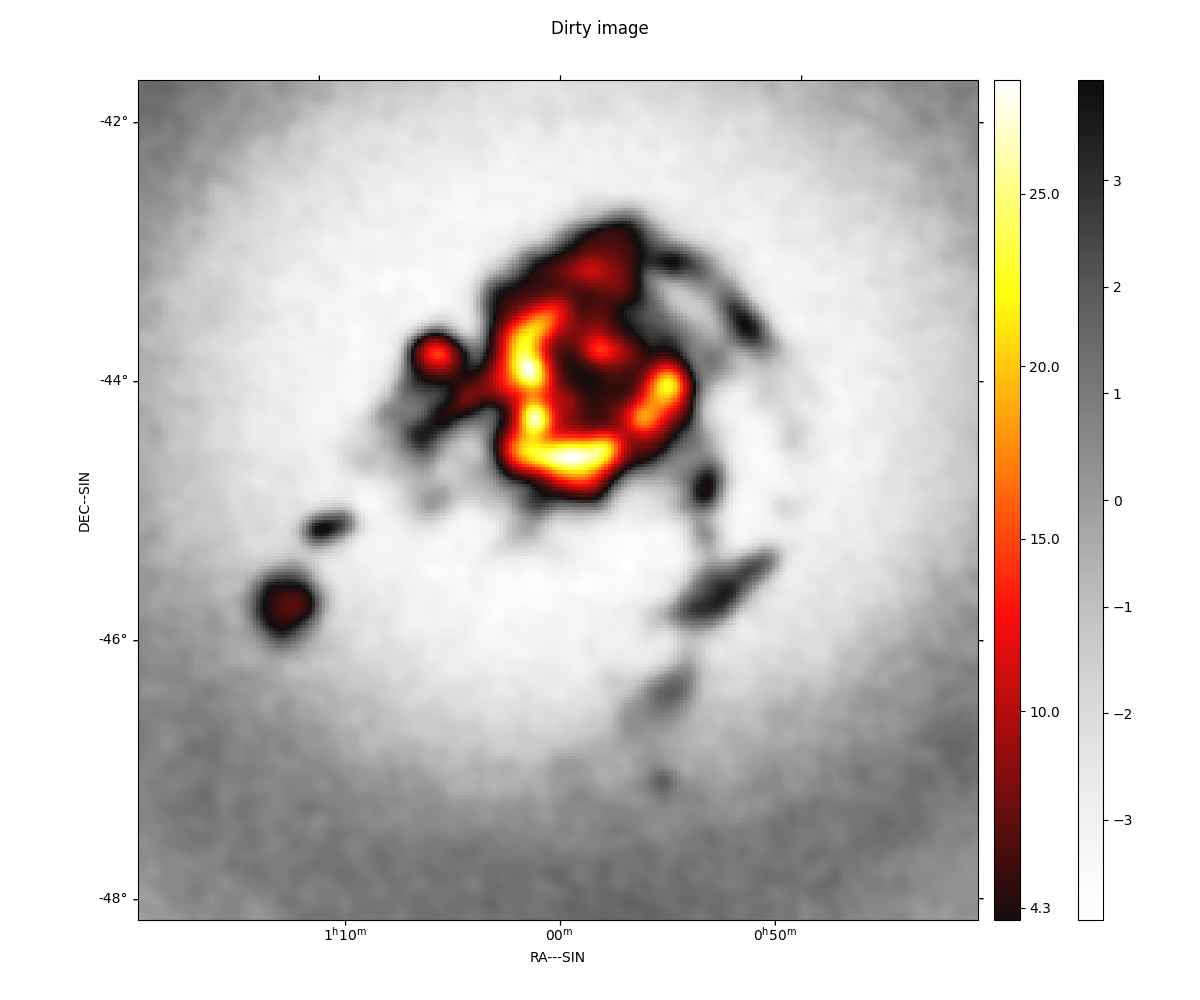}
    \caption{Dirty image obtained with simulated measurements of the Andromeda galaxy (dataset III).}
    \label{fig:ms:dirty}
\end{figure}

\begin{table}[t]
    \centering
    \begin{tabular}{l|cc|cc}
        \toprule
         & \multicolumn{2}{c}{PolyCLEAN} & \multicolumn{2}{c}{\texttt{WSCLEAN}} \\
         \hfill\vadjust{}& MSE & MAD & MSE & MAD \\
         \midrule
         Model images & \textbf{0.1} & \textbf{4.3} & 1.7 & 8.2 \\
         Model convolved  & 12.2 & 45.8 & \textbf{5.7} & \textbf{38.7} \\
         + residuals  & 55.5 & 190.0 & \textbf{14.1} & \textbf{91.4} \\
         \midrule
         Total weight ($\cdot/1495$ Jy)& \multicolumn{2}{c|}{1300} & \multicolumn{2}{c}{1475}\\
         \midrule
         Reconstruction time (s) & \multicolumn{2}{c|}{57.0} & \multicolumn{2}{c}{5.8}\\
         \bottomrule
    \end{tabular}
    \caption{Metrics for reconstruction of diffuse emissions in Sect.~\ref{sec:diffsim} (rescaled, no unit).}
    \label{tab:ms:metrics}
\end{table}

\begin{figure*}
    \centering
    \makebox[\hsize][c]{\includegraphics[width=\textwidth, trim=20 5 0 30, clip]{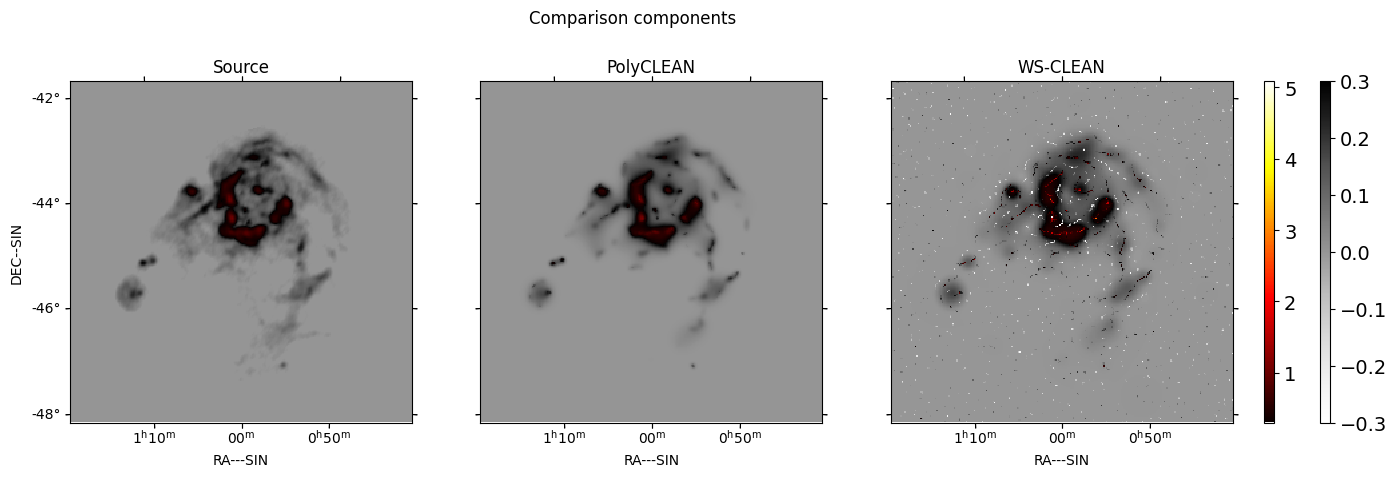}}
    \caption{Model images.}
    \label{fig:ms:components}
\end{figure*}

\begin{figure*}
    \makebox[\hsize][c]{\includegraphics[width=\textwidth, trim=20 5 0 30, clip]{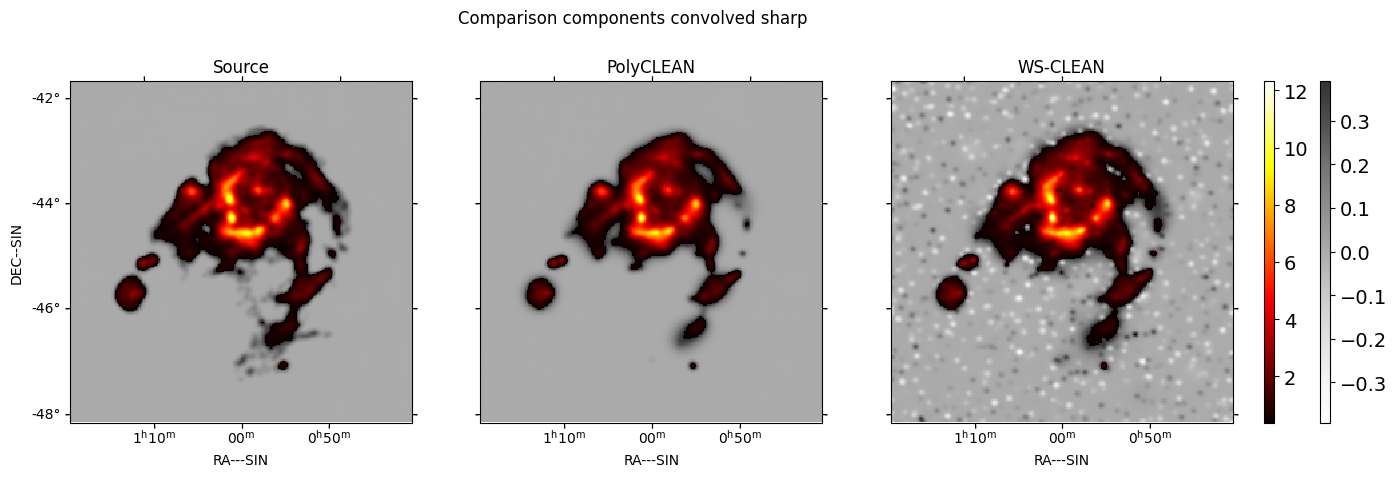}}
    \caption{Model images convolved with a sharp beam.}
    \label{fig:ms:convolved}
\end{figure*}

\begin{figure*}
    \makebox[\hsize][c]{\includegraphics[width=\textwidth, trim=20 5 0 30, clip]{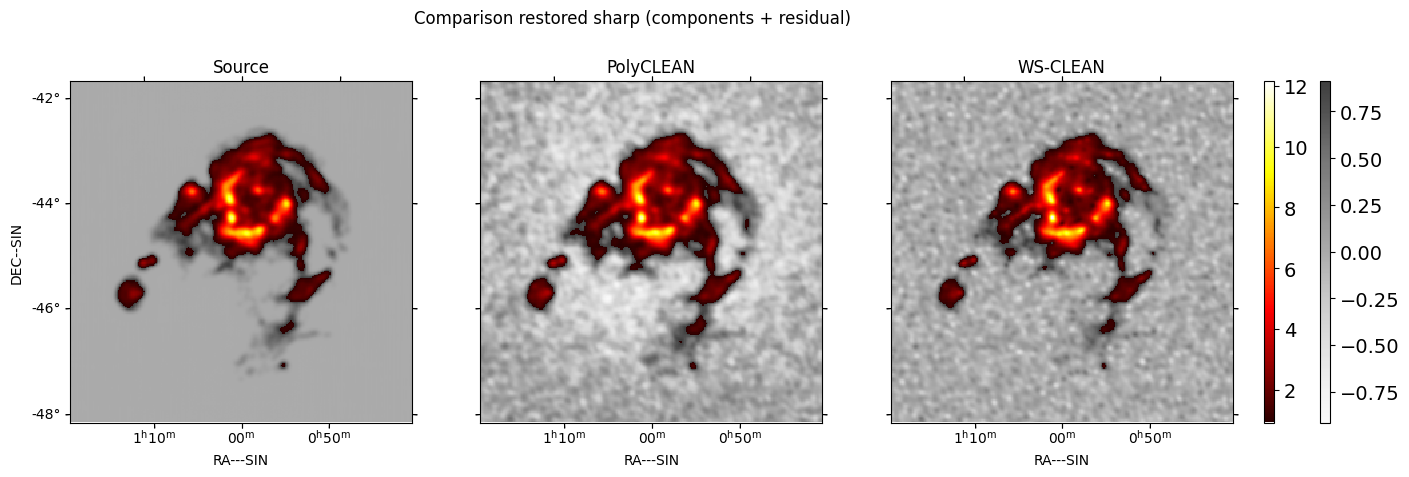}}
    \caption{Reconstructions with residuals.}
    \label{fig:ms:residuals}
\end{figure*}

For the final experiment, we demonstrated that PolyCLEAN can also be used to reconstruct diffuse emissions, making it as versatile and powerful as algorithms from the CLEAN family.

\paragraph{Simulation of the visibilities}
Dataset III consists of a set of simulated measurements from an image of an Halpha region in the Andromeda galaxy (referred to as M31). The source image represents diffuse emissions and is displayed on the left panel of Fig. \ref{fig:ms:components}. We used a test image of size $256\times 256$ provided by \textit{RASCIL} \citep{rascil} and we simulated the visibilities with the SKA-LOW core configuration. The detail of the simulation parameters are reported in Table \ref{tab:datasets}. The test image used here is not meant to be realistic, as the field of view covered by the galaxy on our test image is much larger than its actual size. This choice of simulation parameters allowed us to reduce the size of the baselines so that we could illustrate the reconstruction capabilities of PolyCLEAN on a less demanding setup. The nominal resolution was $113.6$ arcseconds and the image was reconstructed with a slightly finer resolution of $91.4$ arcseconds.

\paragraph{Reconstruction methods}
The two reconstruction methods considered were PolyCLEAN and \texttt{WSCLEAN}, with the following parameters:
\begin{itemize}
    \item PolyCLEAN solved a LASSO problem for sparse dictionary image synthesis following the description in Sect.~\ref{sec:mslasso}. The dictionary was composed of four Gaussian kernels $\mathbf{g}_i$ with different scales $\sigma_i$ and scale-bias factors $\gamma_i$ that were discretized and cropped to a compact support. Formally, we can write $\mathbf{g}_i = \gamma_i \mathbf{\tilde{g}}_i$ with $\mathbf{\tilde{g}}_i [n, m] =(1/G_i) e^{-(n^2 + m^2)/\sigma_i^2}$ for $|n|, |m| \leq 2 \sigma_i$ and $G_i$ a normalization factor such that $\norm{\mathbf{\tilde{g}}_i}_1 = 1$. We considered the scales $\sigma_i = \{0, 2, 5, 8\}$ and the associated weights $\gamma_i = (r)^i$ with $r=1.2$ for $i=\{0, 1, 2, 3\}$ to balance the bias between narrow and wide components in the reconstruction.\footnote{The geometric growth of the scale-bias factors was directly inspired from \cite{Offringa_Smirnov_2017}.} Notably, this multi-scale model allows point sources to be placed with the scale $\sigma_0 = 0$. The regularization parameter was set to $\lambda = 0.02 \lambda_\mathrm{max}$ (note that $\lambda_\mathrm{max}$ then needed to be computed with the synthesis operator $\mathbf{\Psi}$ as well). The algorithm stopped when the relative improvement of the objective function was smaller than $10^{-5}$.

    More generally, the weights $\gamma_i$ of the dictionary elements can be chosen to bias the reconstruction towards using more point sources or more diffuse components. 

    \item \texttt{WSCLEAN} implemented the Multi-scale CLEAN algorithm as described in \cite{Offringa_Smirnov_2017}. In particular, the reconstruction atoms were tapered quadratic functions that are also compactly supported. The \texttt{autothreshold} parameter was again set to one standard deviation.
\end{itemize}

\begin{figure*}
\centering
    \makebox[\hsize][c]{\includegraphics[width=\textwidth, trim=0 5 0 0, clip]{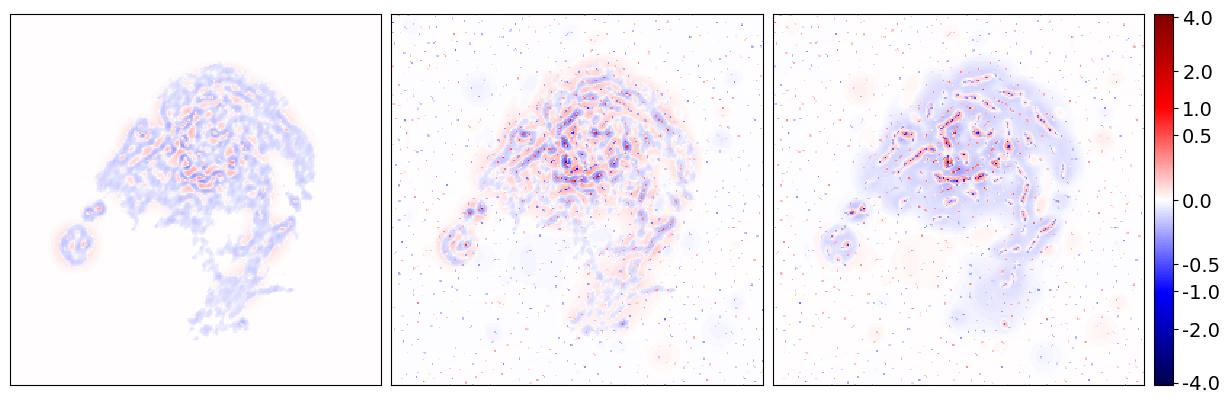}}
    \caption{Difference between model images, square root color scale. \textit{Left:} PolyCLEAN minus source image. \textit{Center:} \texttt{WSCLEAN} minus source image. \textit{Right:} PolyCLEAN minus \texttt{WSCLEAN}.}
    \label{fig:app:diffcomp}
\end{figure*}

\begin{figure*}
    \centering
    \makebox[1.03\hsize][c]{\includegraphics[width=\textwidth, trim=6 5 0 0, clip]{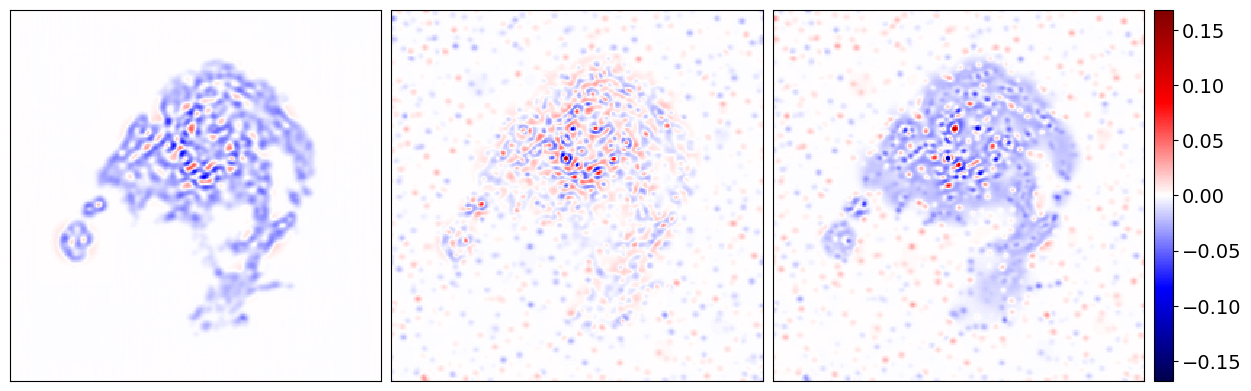}}
    \caption{Difference between convolved images. \textit{Left:} PolyCLEAN minus source image. \textit{Center:} \texttt{WSCLEAN} minus source image . \textit{Right:} PolyCLEAN minus \texttt{WSCLEAN}.}
    \label{fig:app:diffconv}
\end{figure*}

\paragraph{Image reconstruction}
The dirty image is displayed in Fig. \ref{fig:ms:dirty} and the reconstructions are provided in Figs. \ref{fig:ms:components}, \ref{fig:ms:convolved}, \ref{fig:ms:residuals}. These display respectively the model images, the result after convolution with a representation beam, and the final reconstructed images after addition of the residuals. The  model images produced by both algorithms here contain diffuse sources such that these images already represent sources with a spatial extension. For this reason, we used a representation beam more resolved than the CLEAN beam (half of the spread along the two directions). This beam slightly smoothed the models while preserving the recovered multi-scale information. The resulting images are then sharper than if they had been convolved with the CLEAN beam. We also provide some quality metrics in Table~\ref{tab:ms:metrics}. A more detailed view on the center of the images is provided in Appendix~\ref{app:zoomdiff}. Additionally, we compare the reconstructed images with the source and between themselves in Figs.~\ref{fig:app:diffcomp} and \ref{fig:app:diffconv} using difference imaging.

\paragraph{Analysis}
From Figs.~\ref{fig:ms:components} and \ref{fig:ms:convolved}, we see that the two methods reconstructed similar-looking images, both testifying to the multi-scale information of the source. The reconstructions have a similar dynamic range that is consistent with the source. This experiment was run with a pixel size close to the nominal resolution, which partly explains the slow reconstruction speed of PolyCLEAN compared to CLEAN ($10$ times longer).

CLEAN however tended to reconstruct many spurious point sources of small intensity. These are particularly visible on Figs. \ref{fig:ms:components} and \ref{fig:ms:convolved}. This known behavior -- interpreting the noise in the data as additional sources -- lead to the grainy pattern seen in both figures. When the dirty residuals are added in Fig.~\ref{fig:ms:residuals}, the background of the CLEAN image is more regular than that of PolyCLEAN, as it had deliberately left some information aside and identified it as noise. This behavior is also illustrated with the reconstruction metrics provided in Table~\ref{tab:ms:metrics}, in which CLEAN produced better metrics than PolyCLEAN when residuals were considered. Stopping CLEAN earlier would help to alleviate this caveat but at the same time worsen the quality of the recovered emissions.

The PolyCLEAN reconstruction suffered from a loss of overall intensity in the image. We observe this phenomenon at line \textit{Total weight} in Table~\ref{tab:ms:metrics}, which corresponds to the sum of all the reconstructed elements in the model image. This shortcoming of PolyCLEAN is once again explained by the shrinkage effect of LASSO. Despite the fact that PolyCLEAN recovered less information in total, the end solution outperform MS-CLEAN for the model images, demonstrating the relevance of the recovered sources.

Figures~\ref{fig:app:diffcomp} and \ref{fig:app:diffconv} display \textit{difference images} between various reconstructions and so take positive and negative values. Both cases have been normalized by the maximum intensity in the respective source images, hence the displayed value is interpreted as a pointwise error relative to the maximum of the source.

Looking at the differences between model images in Fig.~\ref{fig:app:diffcomp}, we can once again notice the tendency of \texttt{WSCLEAN} to place point sources in the model image. Indeed, on the difference between PolyCLEAN and \texttt{WSCLEAN} (right), sharp blue pixels on the center of the image correspond to the CLEAN sources where PolyCLEAN places smoother red emissions. This creates high values of error so we used a square root scale to enhance the contrast. After convolution with the sharp representation beam (Fig.~\ref{fig:app:diffconv}), the differences between the reconstructions are less intense (colorbar ranges from $-15\%$ to $+15\%$). This convolution is hence of critical importance for CLEAN-based reconstruction but at the same time has a negative impact on the resolution of the recovered image. The PolyCLEAN model image is then an appealing solution to preserve sharpness in the reconstruction.

\section{Conclusions} \label{sec:ccl}

PolyCLEAN proposes the first application of a Frank-Wolfe algorithm to image reconstruction in radio interferometry. By design, it belongs to the family of imaging optimization methods that use the LASSO problem as its guiding reconstruction principle. Despite the model simplicity in comparison with the state-of-the-art in optimization, PolyCLEAN achieved reconstruction quality comparable to modern implementations of CLEAN on both point sources and diffuse emissions. This versatility makes PolyCLEAN an already usable reconstruction framework for actual applications.

The strength of PolyCLEAN lies in its competitive numerical performance in terms of scalability and reconstruction speed. Its design takes full advantage of sparsity properties of the imaging inverse problem involved in radio interferometry. In particular when super-resolution was involved, the increased sparsity in the solutions allowed PolyCLEAN to outperform optimized CLEAN solvers such as \texttt{WSCLEAN}.

PolyCLEAN was developed in Python, with only the computation of the sparse forward operator having been numerically optimized. Further code development is needed to push PolyCLEAN to modern scientific computing standards. Notably, the PolyCLEAN implementation should include parallel distributed computing to tackle the numerical challenges of radio interferometry.

Additionally, we believe that the dual certificate is an insightful tool in the context of radio interferometry. Quantitative imaging techniques will play an essential role in the future of computational imaging, especially with the current rise of learning methods where trust in the reconstruction is of high concern. The dual certificate image makes a step towards an {a posteriori} analysis of the reconstruction.

We finally would like to advocate for the general use of Bayesian methods in computational imaging, as the framework enables data-fidelity and regularization terms used in modern optimization techniques to be interpreted. This results in a better understanding of the reconstruction when compared to CLEAN, for which solution characterization is limited. The Bayesian framework also opens the door for the use of some powerful tools. Extensions considered to PolyCLEAN include the use of Bayesian uncertainty quantification in a similar fashion to \cite{Cai_Pereyra_McEwen_2018b} and hierarchical Bayesian models for determining the hyper-parameters in an automatic, data-driven manner.

\begin{acknowledgements}
The authors sincerely thank Martin Vetterli for his guidance and support throughout the writing of this article. We also  thank the team at the EPFL Center for Imaging for help on computational aspects and use of their server. We finally acknowledge the anonymous referee whose review contributed to significantly improve the quality of this document.

A.J. is funded by the Swiss National Science Fundation (SNSF) under grants \textit{AstroSignals: A New Window on the Universe, with the New Generation of Large Radio-Astronomy Facilities, Sinergia (CRSII5\_193826)} and \textit{SESAM - Sensing and Sampling: Theory and Algorithms (n° 200021\_181978/1)}.
\end{acknowledgements}

\bibliographystyle{bibtex/aa}
\bibliography{bibtex/refs.bib}

\begin{appendix}
\onecolumn
\section{Scaling of the simulated problems}
\label{app:pbm-size}

\begin{table}[h]
    \caption{Dimensions of the simulated problems in the  experiment from Sect.~\ref{sec:profile}. For each value of $r_\mathrm{max}$, the number of visibilities considered increase and we have access to higher frequencies. Hence, more pixels can be used to cover the same field of view and the resolution of the image increases. This is valid for any chosen super resolution factor. }
    \label{tab:scaling}
    \centering
    \begin{tabular}{l|r|rr|rr|rr}
        \multirow{2}{*}{$r_\mathrm{max}$ (km)} & Number of & \multicolumn{2}{c}{SRF 2} & \multicolumn{2}{c}{SRF 5} & \multicolumn{2}{c}{SRF 10} \\
         & visibilities ($10^3$) & Pix/side & Size (MPix) & Pix/side & Size (MPix) & Pix/side & Size (MPix) \\
        \midrule
         0.3 &    30.6 &  144 &  0.02 &  360 &  0.13 &   720 &     0.5 \\
         0.6 &   168.6 &  288 &  0.08 &  720 &  0.52 &  1440 &     2.1 \\
         0.9 &   209.2 &  384 &  0.15 &  960 &  0.92 &  1920 &     3.7 \\
         1.2 &   235.7 &  432 &  0.19 & 1080 &  1.17 &  2160 &     4.7 \\
         1.5 &   269.5 &  576 &  0.33 & 1440 &  2.07 &  2880 &     8.3 \\
         2.0 &   293.3 &  720 &  0.52 & 1800 &  3.24 &  3600 &    13.0 \\
         3.0 &   344.0 & 1080 &  1.17 & 2700 &  7.29 &  5400 &    29.2 \\
         6.0 &   457.4 & 2000 &  4.00 & 5000 & 25.00 & 10000 &   100.0 \\
    \end{tabular}
\end{table}

\section{Error metrics}
\label{app:metrics}

\begin{figure}[h]
    \resizebox{\hsize}{!}{\includegraphics[trim=100 0 100 0, clip]{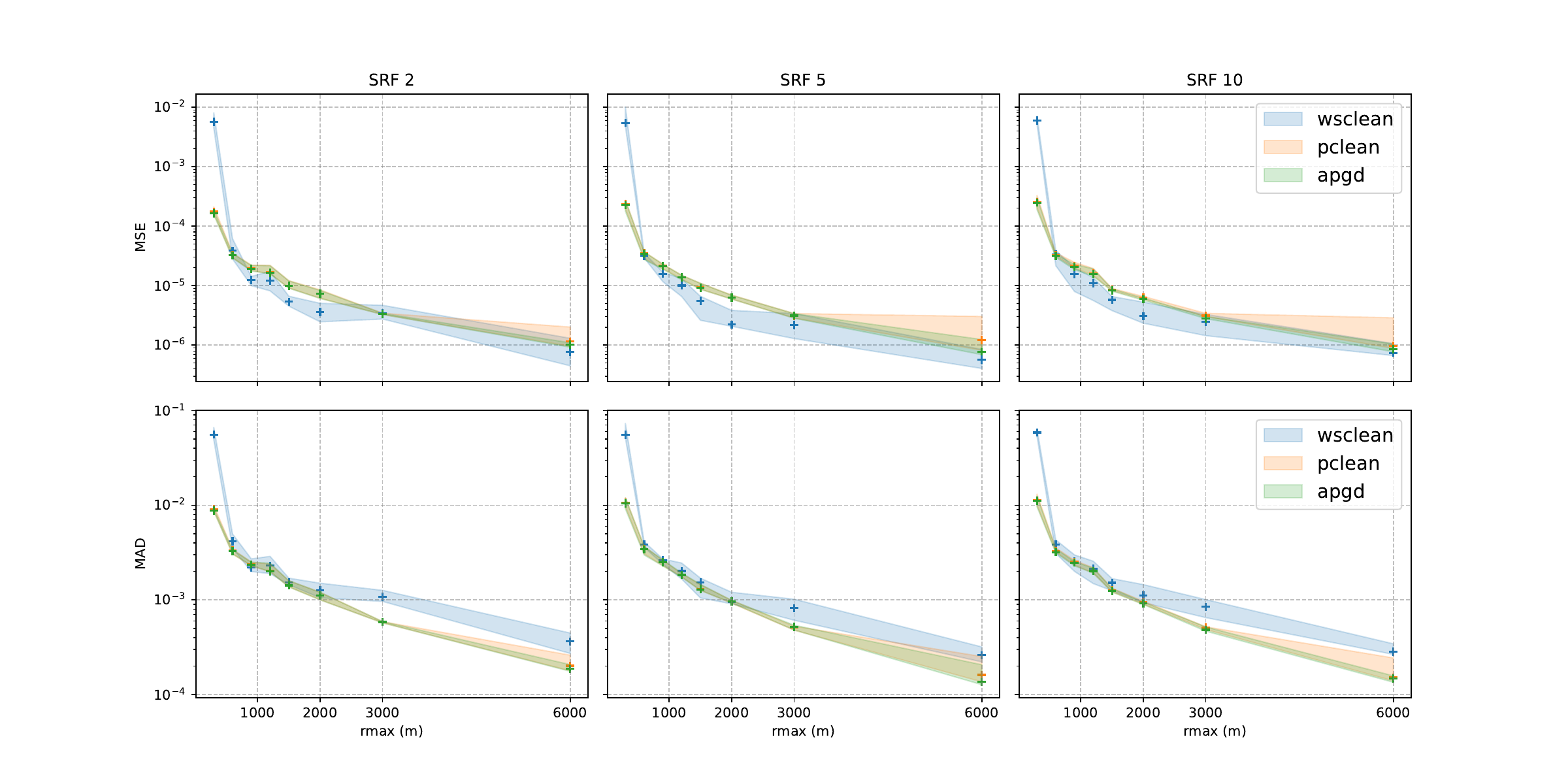}}
    \caption{Comparative view of the reconstruction errors on the simulated images during the experiment of Sect~\ref{sec:profile}. Median values over the 10 repetitions of the experiments are reported with the crosses and the transparent areas cover the interquartile spread. MSE appears on the first row, MAD on the second one. The metrics are computed after convolution with the CLEAN beam. Each column is a different super resolution factor: 2, 5, 10 (from left to right). We observe that the metrics have similar values between CLEAN and the LASSO-based methods (PolyCLEAN and APGD). This indicates that the different algorithms are indeed configured to return images of similar quality and thus the comparison between the solving times performed in the experiment is relevant.}
     \label{fig:msemad}
\end{figure}

\clearpage

\section{Reconstruction examples}
\label{app:reco}

\begin{figure}[h]
    \begin{subfigure}{\hsize}
        \centering
        \includegraphics[width=\hsize, trim=5 0 5 0, clip]{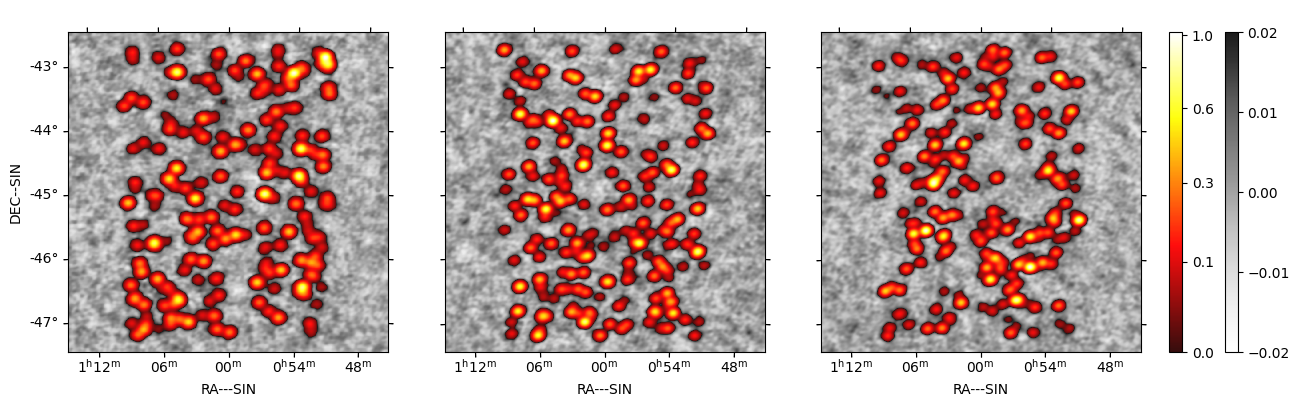}
    \end{subfigure}

    \medskip

    \begin{subfigure}{\hsize}
        \centering
        \includegraphics[width=\hsize, trim=5 0 5 0, clip]{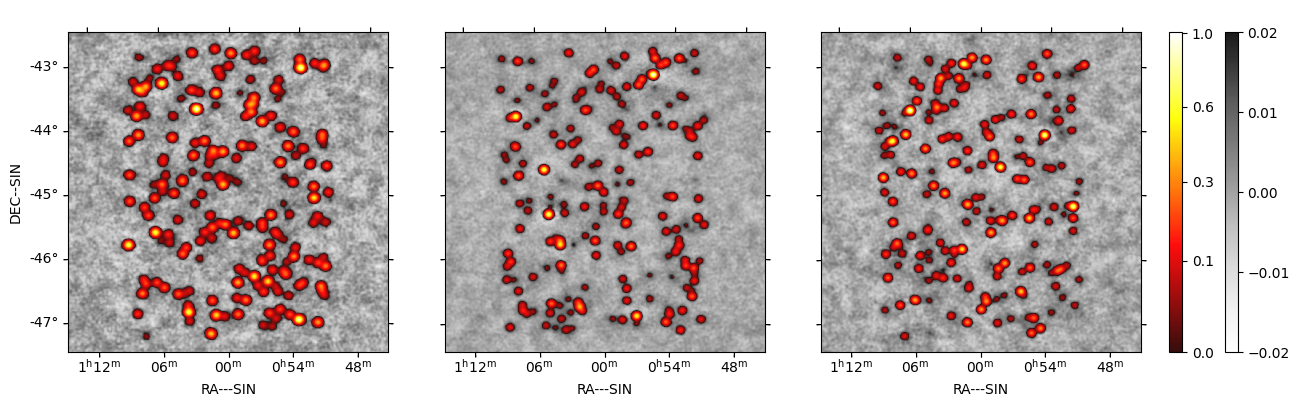}
    \end{subfigure}
    
    \medskip

    \begin{subfigure}{\hsize}
        \centering
        \includegraphics[width=\hsize, trim=5 0 5 0, clip]{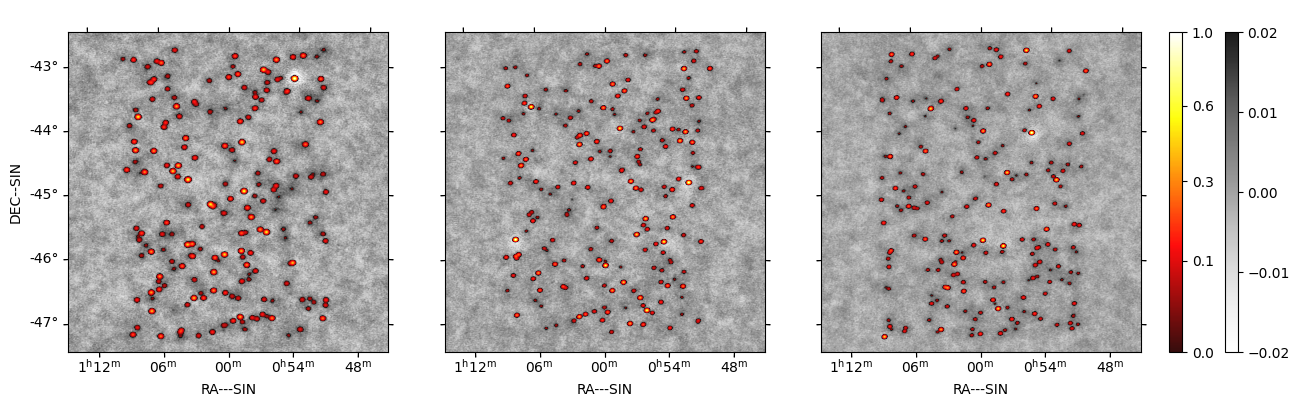}
    \end{subfigure}
    
    \caption{PolyCLEAN reconstructions of simulated skies in the experiment of Sect~\ref{sec:profile}. Each row uses a different $r_\mathrm{max}$ and each column is with a different SRF. From top to bottom: $2000$, $3000$, $6000$ -- from left to right: \textit{SRF2}, \textit{SRF5}, \textit{SRF10}. The field of view is kept fixed but the number of pixels increases from left to right and from top to bottom. As a consequence of including longer baselines, the CLEAN beam also gets narrower (from top to bottom).}
    \label{fig:app:reco}
\end{figure}

\clearpage

\section{Difference imaging on observations}
\label{app:diff-obs}

\begin{figure}[h]
    \centering
    \resizebox{0.89\hsize}{!}{
    \includegraphics[trim=20 18 20 45, clip]{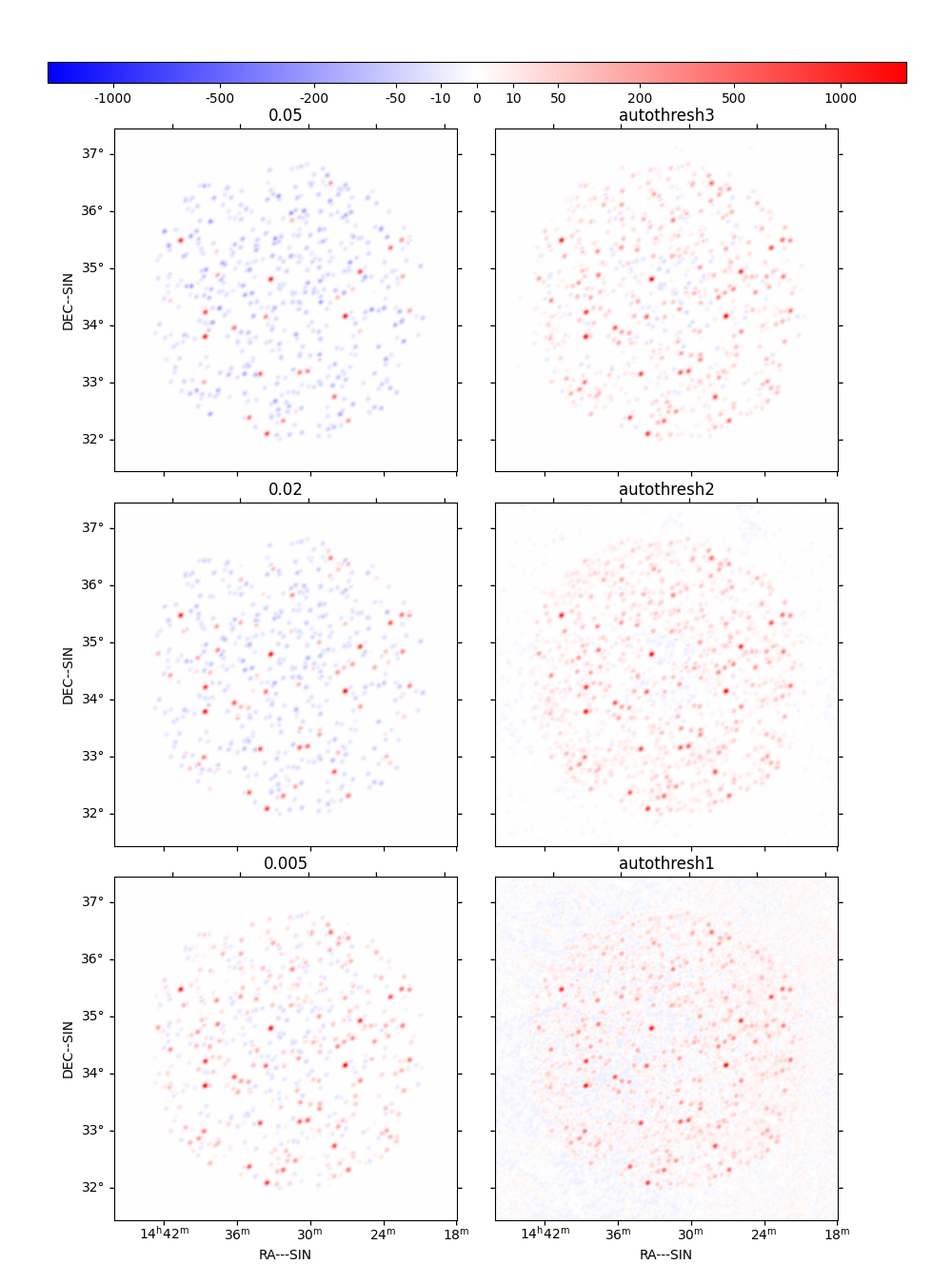}}
    \caption{Differences between the reconstructions and the catalog image from \citep{williams2016}. Red and blue respectively indicate over- and under-estimation compared to the catalog. Intensities are displayed with a square root colorscale.}
    \label{fig:app:differences}
\end{figure}

We observe that some strong sources are overestimated both with PolyCLEAN and \texttt{WSCLEAN} (bright red dots). This consistent behavior between the two methods probably originates from a mismatch between our observation data and the catalog. Location errors appear as a red dot next to a blue one but very few are present here. Additionally, we once again observe the shrinkage behavior of the LASSO problem with the PolyCLEAN reconstructions as well the tendency of CLEAN to place sources to explain the noise when the algorithm runs too long.

\section{Details on reconstruction of diffuse emissions}
\label{app:zoomdiff}

\begin{figure}[h]
    \centering
    \makebox[\hsize][c]{\includegraphics[width=\textwidth, trim=20 5 0 30, clip]{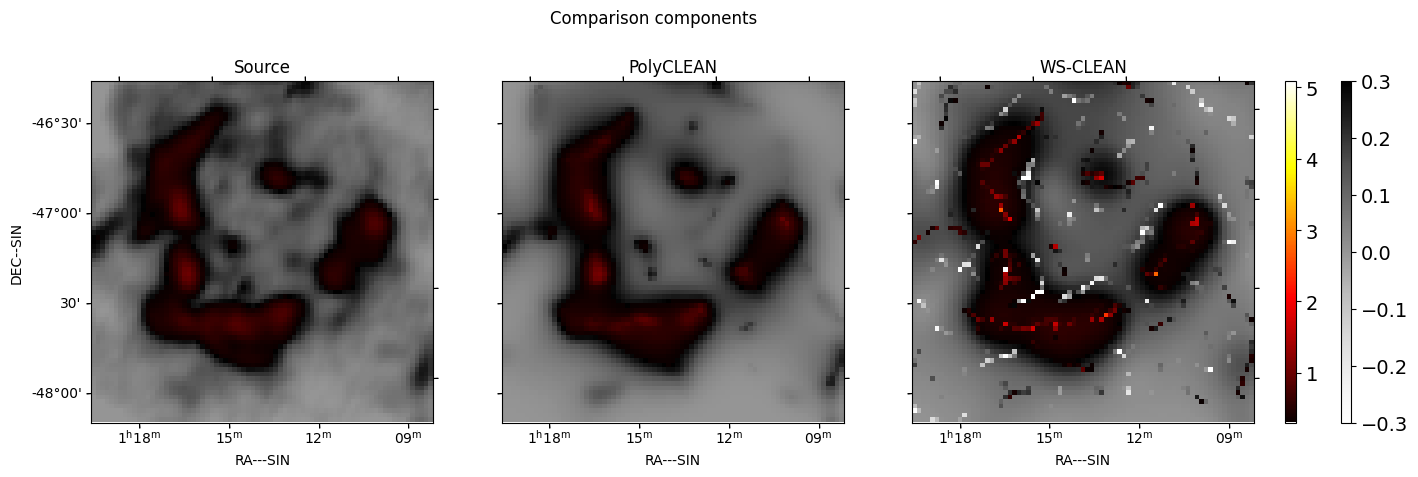}}
    \caption{Model images.}
    \label{fig:app:compzoom}
\end{figure}

\begin{figure}[h]
    \centering
    \makebox[\hsize][c]{\includegraphics[width=\textwidth, trim=20 5 0 30, clip]{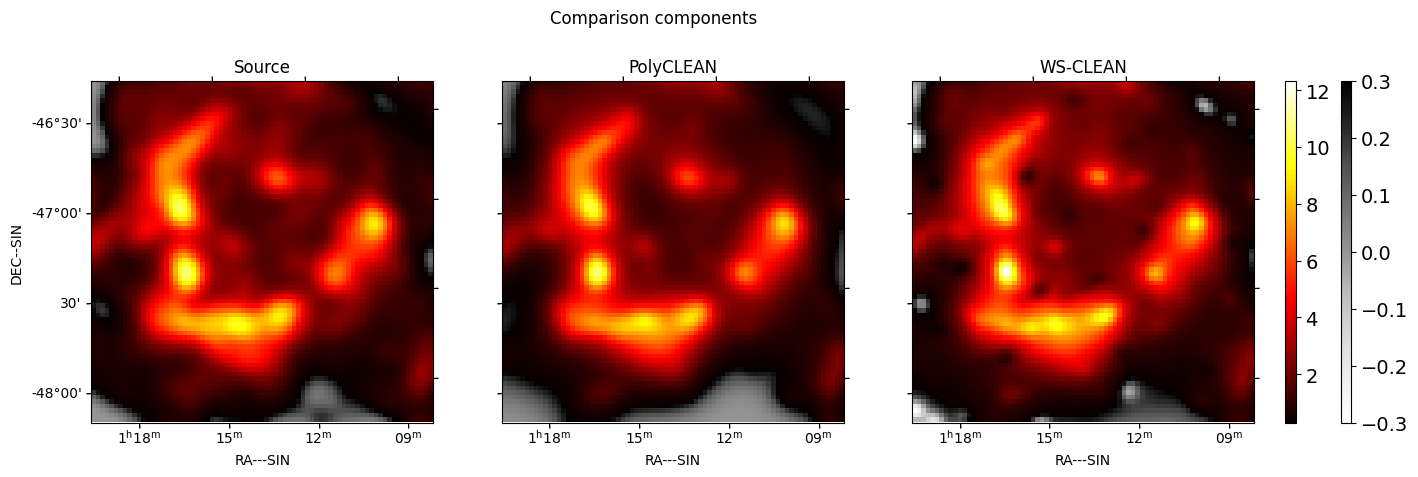}}
    \caption{Convolution with a sharp beam.}
    \label{fig:app:convzoom}
\end{figure}

\begin{figure}[h]
    \centering
    \makebox[\hsize][c]{\includegraphics[width=\textwidth, trim=20 5 0 30, clip]{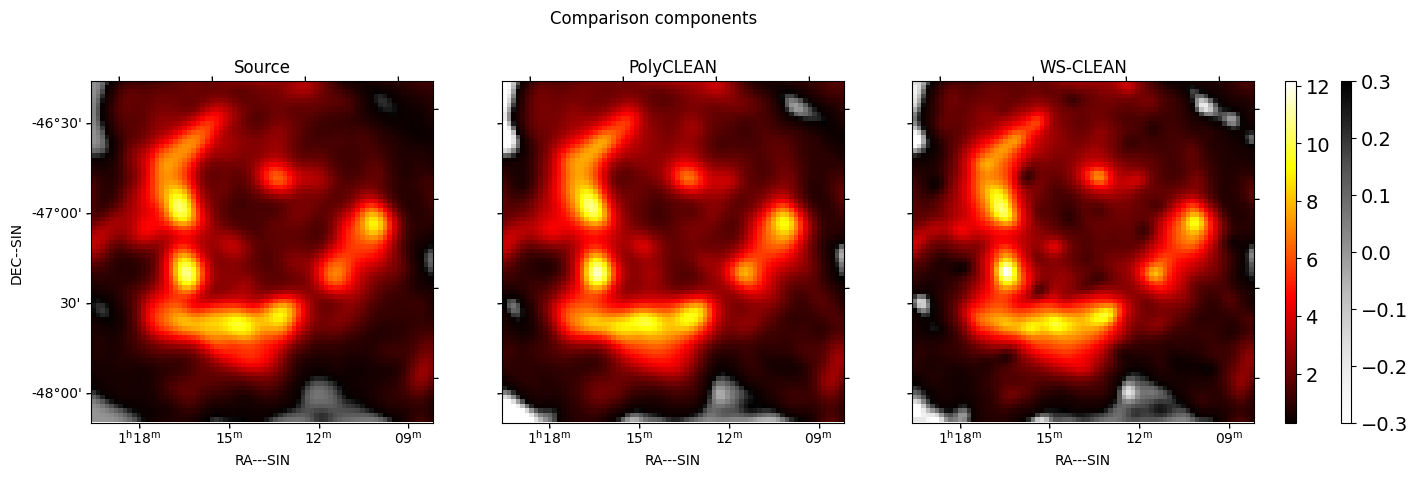}}
    \caption{Reconstructions with residuals.}
    \label{fig:app:restoredzoom}
\end{figure}

We observe that \texttt{WSCLEAN} produced many single-pixel point sources in the reconstruction, even though it was configured to recover diffuse emissions, as we clearly see in Fig.~\ref{fig:app:compzoom}. The model image proposed by PolyCLEAN involves more diffuse components, which make the image visually closer to the source. It resulted in better metrics, as reported in Table~\ref{tab:ms:metrics}.

After convolution however, the difference between the two reconstructions is more difficult to visually analyze (Fig.~\ref{fig:app:convzoom}). The hotspots displayed the same intensity and had the same locations, which are consistent with the source image. This representation smoothed out the singularities introduced by the tendency of \texttt{WSCLEAN} to place point sources. Note that the use of a sharp representation kernel here is not conventional and common practice consists in using the CLEAN beam, spreading more evenly the information and hence loosing precision. CLEAN artifacts are still noticeable with the presence of spurious bright negative areas, even after convolution.

For completeness, we also provide the images after adding the dirty residual in Fig.~\ref{fig:app:restoredzoom} (except on the source image). Interestingly, we observe that the reconstructed images (\textit{i.e.}, PolyCLEAN and \texttt{WSCLEAN}) seem to be visually closer to the convolved source in some areas. We can interpret this as indicating that the residuals still contained some relevant information.

\end{appendix}

\end{document}